\documentclass[iop]{emulateapj-rtx4}
\usepackage{apjfonts}

\newcommand{\B}{$m_{\rm F438W}$}
\newcommand{\V}{$m_{\rm F555W}$}

\newcommand{\BI}{$m_{\rm F438W} - m_{\rm F814W}$}

\newcommand{\VI}{$m_{\rm F555W} - m_{\rm F814W}$}
\newcommand{\Msun}{M$_{\odot}$}
\newcommand{\kms}{km\,s$^{-1}$} 

\slugcomment{Accepted for publication in ApJ}

\shorttitle{On the nature of eMSTOs in LMC Star Clusters}
\shortauthors{Correnti et al.}

\begin{document}


\title{New Clues to the Cause of Extended Main Sequence Turn-Offs in
  Intermediate-Age Star Clusters in the  Magellanic Clouds\altaffilmark{1}} 


\author{Matteo Correnti\altaffilmark{2}, Paul Goudfrooij\altaffilmark{2}, 
Jason S. Kalirai\altaffilmark{2,3}, Leo Girardi\altaffilmark{4}, Thomas H. Puzia\altaffilmark{5}, and Leandro Kerber\altaffilmark{6}}

\altaffiltext{1}{Based on observations with the NASA/ESA {\it Hubble Space Telescope}, obtained at the Space Telescope Science Institute, which is operated by the Association of Universities for Research in Astronomy, Inc., under NASA contract NAS5-26555}

\altaffiltext{2}{Space Telescope Science Institute, 3700 San Martin Drive, Baltimore, MD 21218, USA; correnti, goudfroo, jkalirai@stsci.edu}

\altaffiltext{3}{Center for Astrophysical Science, Johns Hopkins University, Baltimore, MD 21218, USA}

\altaffiltext{4}{Osservatorio Astronomico di Padova, INAF, Vicolo dell'Osservatorio 5, 35122, Padova, Italy; leo.girardi@oapd.inaf.it}

\altaffiltext{5}{Institute of Astrophysics, Pontificia
Universidad Cat\'olica de Chile, Av.\ Vicu\~{n}a Mackenna 4860, Macul
7820436, Santiago, Chile; tpuzia@astro.puc.cl} 

\altaffiltext{6}{Universidade Estadual de Santa Cruz, Rodovia Ilh\'eus-Itabuna, km 16, 45662-000 Ilh\'eus, Bahia, Brazil; lkerber@gmail.com}

\begin{abstract}
We use the Wide Field Camera 3 onboard the Hubble Space Telescope (HST) to
obtain deep, high resolution images of two intermediate-age star clusters in the
Large Magellanic Cloud of relatively low mass ($\approx 10^4\; M_{\odot}$) and
significantly different core radii, namely NGC~2209 and NGC~2249. For comparison
purposes, we also re-analyzed archival HST images of NGC~1795 and IC~2146, two
other  relatively low mass star clusters. From the comparison of the observed
color-magnitude diagrams with Monte Carlo simulations, we find that the main
sequence turnoff (MSTO) regions in NGC~2209 and NGC~2249 are significantly wider
than  that derived from simulations of simple stellar populations, while those
in NGC~1795 and IC~2146 are not. We determine the evolution of the clusters'
masses and escape velocities from an age of 10 Myr to the present age. We find
that differences among these clusters can be explained by dynamical evolution
arguments if the currently extended clusters (NGC~2209 and IC~2146) experienced
stronger levels of initial mass segregation than the currently compact ones
(NGC~2249 and NGC~1795). Under this assumption, we find that NGC~2209 and
NGC~2249 have estimated escape velocities $V_{\rm esc} \ga 15$ \kms\ at an age
of 10 Myr, large enough to retain material ejected by slow winds of
first-generation stars, while the two clusters that do \emph{not} feature
extended MSTOs have $V_{\rm esc} \la 12$ \kms\ at that age. These results
suggest that the extended MSTO phenomenon can be better explained by a range of
stellar ages rather than a range of stellar rotation velocities or interacting
binaries. 
\end{abstract}




\section{Introduction}
Recently, deep Color-Magnitude Diagrams (CMDs) from images taken with the
Advanced Camera for Survey (ACS) aboard the Hubble Space Telescope (HST)
revealed that several massive intermediate-age ($\sim$\,1\,--\,2 Gyr old) star
clusters in the Magellanic Clouds host extended and/or multiple main sequence
turn-offs (MSTOs) regions (\citealt{mack+08a,glat+08,milo+09,goud+09};
\citealt{goud+11a}, hereafter G11a; \citealt{piat13}), in some cases
accompanied by composite red clumps (\citealt{gira+09,rube+11}). A popular
interpretation of these extended MSTOs (herafter eMSTOs) is that they are due to
stars  that formed at different times within the parent cluster, with an age
spread of 150\,--\,500 Myr (\citealt{milo+09,gira+09,rube+10,rube+11}; G11a).
Other potential causes of eMSTOs mentioned in the recent literature include
spreads in rotation velocity among turn-off stars (\citealt{basdem09,yang+13},
but see \citealt{gira+11}), a photometric feature of interacting binary stars
(\citealt{yang+11}), or a combination of both (\citealt{li+12}). 

A relevant aspect of the nature of the eMSTO phenomenon among intermediate-age
star clusters is that it is \emph{not} shared by all such clusters (e.g.,
\citealt{milo+09}, G11a). In this context, two scenarios have been proposed in
the recent literature to predict the existence of eMSTOs in intermediate-age
clusters. \citet[][hereafter K11]{kell+11} noted that the eMSTOs known at the
time were all hosted by intermediate-age clusters that have a large core radius
($r_c \ga 3.7$ pc), and suggested that such a large core radius is a
prerequisite for hosting an eMSTO. Their argument is based on simulations of
star cluster formation indicating that the mass loss due to stellar evolution in
a primordially mass-segregated cluster can lead to a significant cluster core
expansion and low central density \citep{mack+08b}. Since mass segregation is
also expected to be a function of the cluster mass \citep[e.g.,][]{bonbat06},
K11 suggested that the clusters with the largest core radii represent the
initially most mass-segregated (and hence likely most massive) clusters to have
formed. In a similar but not identical argument, \citet[][hereafter
G11b]{goud+11b} suggest that the key factor to explain the presence of
intermediate-age clusters hosting an eMSTO is the cluster's ability to retain
the material ejected by first-generation stars that are thought to be
responsible for the formation of a second generation (often called
``polluters''). Following the  arguments of G11b, eMSTOs can be formed only if
the cluster escape velocity was higher than the wind velocity of such polluter
stars when the latter existed. 

Currently, the most popular candidates for first-generation ``polluters'' are
{\it (i)\/} intermediate-mass AGB stars ($4 \la {\cal{M}}/M_{\odot} \la 8$,
hereafter IM-AGB; e.g., \citealt{danven07}, and references therein), {\it
(ii)\/} rapidly rotating massive stars (often referred to as ``FRMS''; e.g.
\citealt{decr+07}) and {\it (iii)\/} massive binary stars \citep{demink+09}.
Wind velocities of these stars must be compared with the cluster escape velocity
derived at the same time these stars were present in the cluster (i.e. at ages
of $\sim$ 5\,--\,30 Myr for massive stars and $\sim$ 50\,--\,200 Myr for IM-AGB
stars). 

We note that massive ($\ga 10^5\;M_{\odot}$) intermediate-age star clusters do
not provide very strong constraints to the cause of the presence of eMSTOs. This
is due to the fact that  most of them contain large core radii which renders
both scenarios to predict the presence of eMSTOs in those clusters. On the other
hand, \emph{low-mass}  (i.e. $\approx 10^4\; M_{\odot}$) star clusters
\emph{can} provide important insights into the nature of eMSTOs, especially when
selecting  such clusters with a variety of core radii. With this in mind, we
present the analysis of new two-color HST Wide Field Camera 3 (WFC3) photometry
of two intermediate-age star clusters in the Large Magellanic Cloud (LMC) of
relatively low mass and significantly different radii, namely NGC~2209 and
NGC~2249. \cite{kell+12} used ground-based Gemini/GMOS photometry to study
NGC~2209; the authors identified an eMSTO, but the limited spatial resolution
prevented them from a detailed analysis of the MSTO morphology in the inner
regions of the cluster. For comparison purposes, we also re-analyzed archival
HST/ACS images of two other relatively low mass star clusters, namely NGC~1795
and IC~2146, for which \cite{milo+09} already showed that they do not exhibit an
eMSTO. We conduct a detailed investigation of the MSTO morphology of the four
clusters, comparing the observed CMDs with Monte Carlo simulations in order to
quantify the widths of MSTO regions and verify whether they can be reproduced by
a simple stellar population (SSP). Moreover, we study the evolution of the
clusters' masses and escape velocities from an age of 10 Myr to their current
age. This analysis allows us to reveal new findings relevant to the causes for
the formation and retention of second-generation stars in these intermediate-age
star clusters.
 
The remainder of this paper is organized as follows: Section~2 presents the
clusters' observations and data reduction. In Section~3 we present CMDs of the
four clusters and the comparison with Monte Carlo simulations. In Section~4 we
derive the evolution of the clusters' masses and escape velocities as a function
of age, using models with and without initial mass segregation. Finally, in
Section~5 we present and discuss our main results.  

\section{Observations and Data Reduction}
\label{s:obs}
NGC~2209 and NGC~2249 were observed with HST on 2013 May 23 and 2013 November
11, respectively, using the UVIS channel of the WFC3 as part of HST program
12908 (PI: P.\ Goudfrooij). The clusters were centered on one of the two CCD
chips of the WFC3 UVIS camera, so that the observations cover enough radial
extent to study variations with cluster radius and to avoid the loss of the
central region of the clusters due to the CCD chip gap. Two long exposures were
taken in each of the F438W and F814W filters: for NGC~2209, their exposure times
were 850 s (F438W filter) and 485 s (F814W filter). For NGC~2249, they were 825
s and 425 s, respectively. In addition, we took one short 60 s exposure in F814W
for each cluster in order to avoid saturation of the brightest RGB and AGB
stars.  The two long exposures in each filter were spatially offset from each
other  by 2\farcs401 in a direction +85\fdg76 with respect to the positive
X-axis of the CCD array. This was done to move across the gap between the two
WFC3/UVIS CCD chips, as well to simplify the identification and removal of hot
pixels and cosmic rays. In addition to the WFC3/UVIS observations, we used the
Wide Field Camera (WFC) of ACS in parallel to obtain images $\approx 6'$ from
the cluster centers. These images have been taken in the same filters used for
the prime observations of the cluster (i.e., F435W and F814W). The ACS exposure
times were adjusted in order to hide the parallel buffer dumps without
sacrificing the target S/N of the WFC3 images. These ACS images provide valuable
information of the stellar content and Star Formation History (SFH) in the
underlying LMC field, permitting us to establish in detail the field  star
contamination fraction in each region of the CMDs. They also allow a valuable
measurement of the ``background'' stellar surface density in the context of our
King model fits (see Section~\ref{s:dynamics}). For NGC~1795 and IC~2146, we
retrieved the STScI MAST archive images collected with ACS/WFC as part of the
HST program GO-9891 (PI: G.\ F.\ Gilmore). Descriptions of those observations
and datasets is reported in Table~1 of \cite{milo+09}.

The reduction of the images of NGC~2209 and NGC~2249 was performed following the
method described in  \citet{kali+12}. Briefly, we started from the {\it flt}
files provided by the HST pipeline. The {\it flt} files constitute the
bias-corrected, dark-subtracted and flat-fielded images. We then corrected all
the {\it flt} files for charge transfer inefficiency, using the CTE correction
software\footnote{http://www.stsci.edu/hst/wfc3/tools/cte\_tools}. We generated
distortion-free images using MultiDrizzle \citep{fruchook97} and we calculated
the transformations between the individually drizzled images to link them to a
reference frame (i.e.:\ the first exposure) in each filter. The transformations
were based on  Gaussian-fitted centroids of hundreds of stars on each image and
the solution was refined through successive matches. The final transformations
provide alignment of the individual images to better than 0.02 pixels. These
offsets were then supplied to a final run of MultiDrizzle as a ``shift'' file
and added to the WCS header information. Finally, bad pixels and cosmic rays are
flagged and rejected from the input images, and a final image is created for
each filter, combining the input undistorted and aligned frames.  The final
stacked images were generated at the native resolution of the WFC3/UVIS and
ACS/WFC (i.e.: 0\farcs040 pixel$^{-1}$ and 0\farcs049 pixel$^{-1}$,
respectively). For NGC~1795 and IC~2146, we used the {\it drc} files produced by
the HST/ACS pipeline. 

Stellar photometry was performed on the stacked images, using the stand-alone
versions of the DAOPHOT-II and ALLSTAR point spread function (PSF) fitting
programs \citep{stet87,stet94}. The final catalog is based on first performing
aperture photometry on all the sources that are at least 3$\sigma$ above the
local sky, then deriving a PSF from $\sim 1000$ bright isolated stars in the
field, and finally applying the PSF to all the sources detected in the aperture
photometry list. The final catalogs contain sources that were iteratively
matched between the two images, and have been cleaned by eliminating background
galaxies and spurious detections by means of $\chi^2$ and sharpness cuts from
the PSF fitting. 

To perform the photometric calibration, we used a sample of bright isolated
stars to transform the instrumental PSF-fitted magnitudes to a fixed aperture of
10 pixels (0\farcs40 for WFC3/UVIS, 0\farcs49 for ACS/WFC). The magnitudes were
then transformed into the VEGAMAG system by adopting the relevant synthetic zero
points for the WFC3/UVIS and ACS/WFC bands. 

To quantify incompleteness, we used the standard technique of adding artificial
stars to the images and running them through the photometric pipeline. A total
of nearly 120000 artificial stars were added to each image. In order to leave
the crowding conditions unaltered, only $\sim 5\%$ of the total number of stars
in the final catalogs were added per run; the overall distribution of the
inserted artificial stars followed that of the stars in the image. The
artificial stars were distributed in magnitude according to a luminosity
function similar to the observed one and with a color distribution covering the
full color ranges found in the CMDs. After inserting the artificial stars, the
photometry procedures described above were applied again to the image. An
inserted star was considered recovered if the input and output magnitudes agreed
to within 0.75 mag in both filters. Completeness fractions were assigned to
every individual star in a given CMD by fitting the completeness fractions as a
function of the magnitude and distance from the cluster center.  

\section{Color-Magnitude Diagrams and Monte Carlo Simulations} 
\label{s:photometry}
\begin{figure*}[pt]
\includegraphics[scale=0.29]{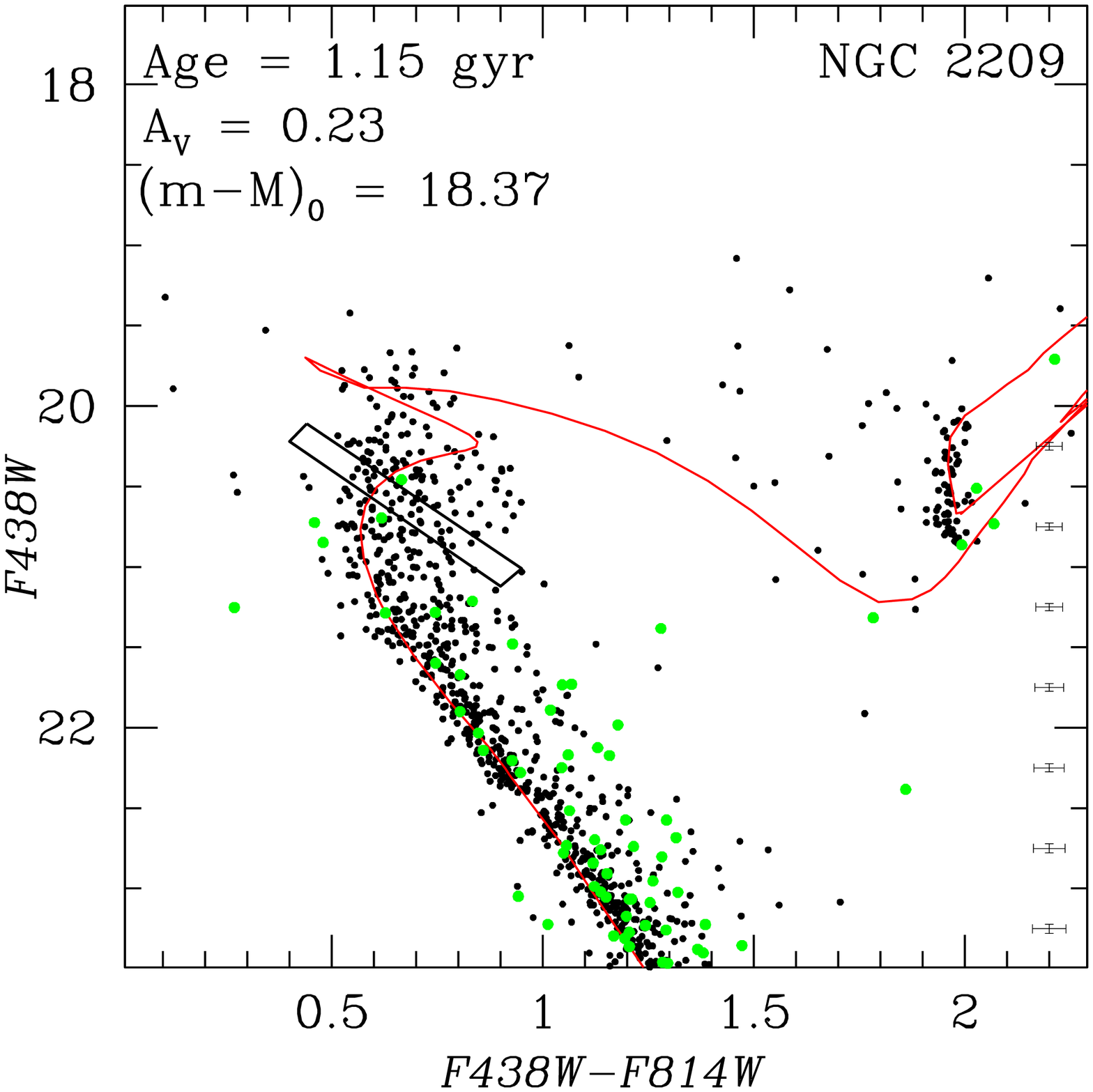}
\includegraphics[scale=0.29]{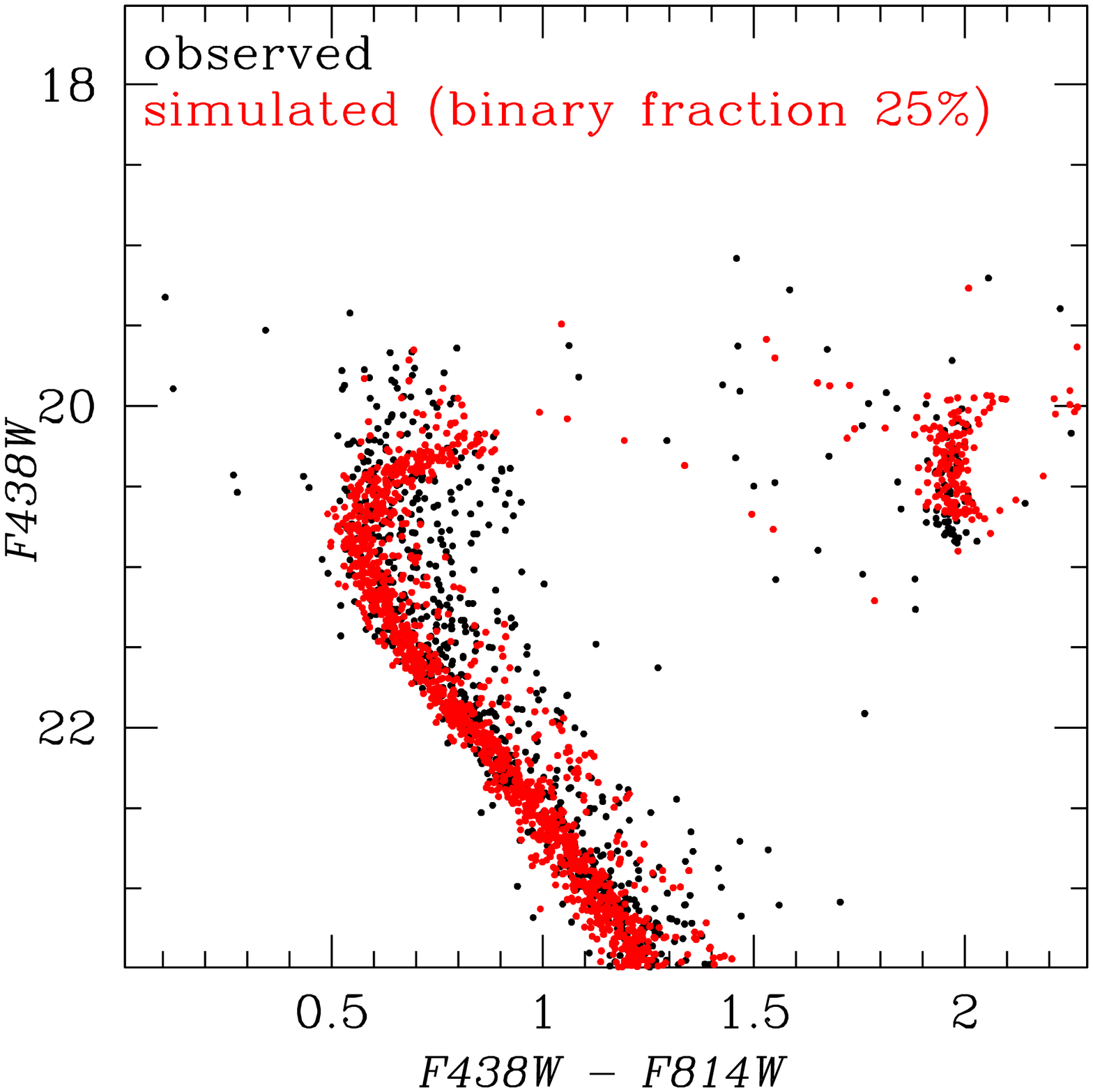}
\includegraphics[scale=0.29]{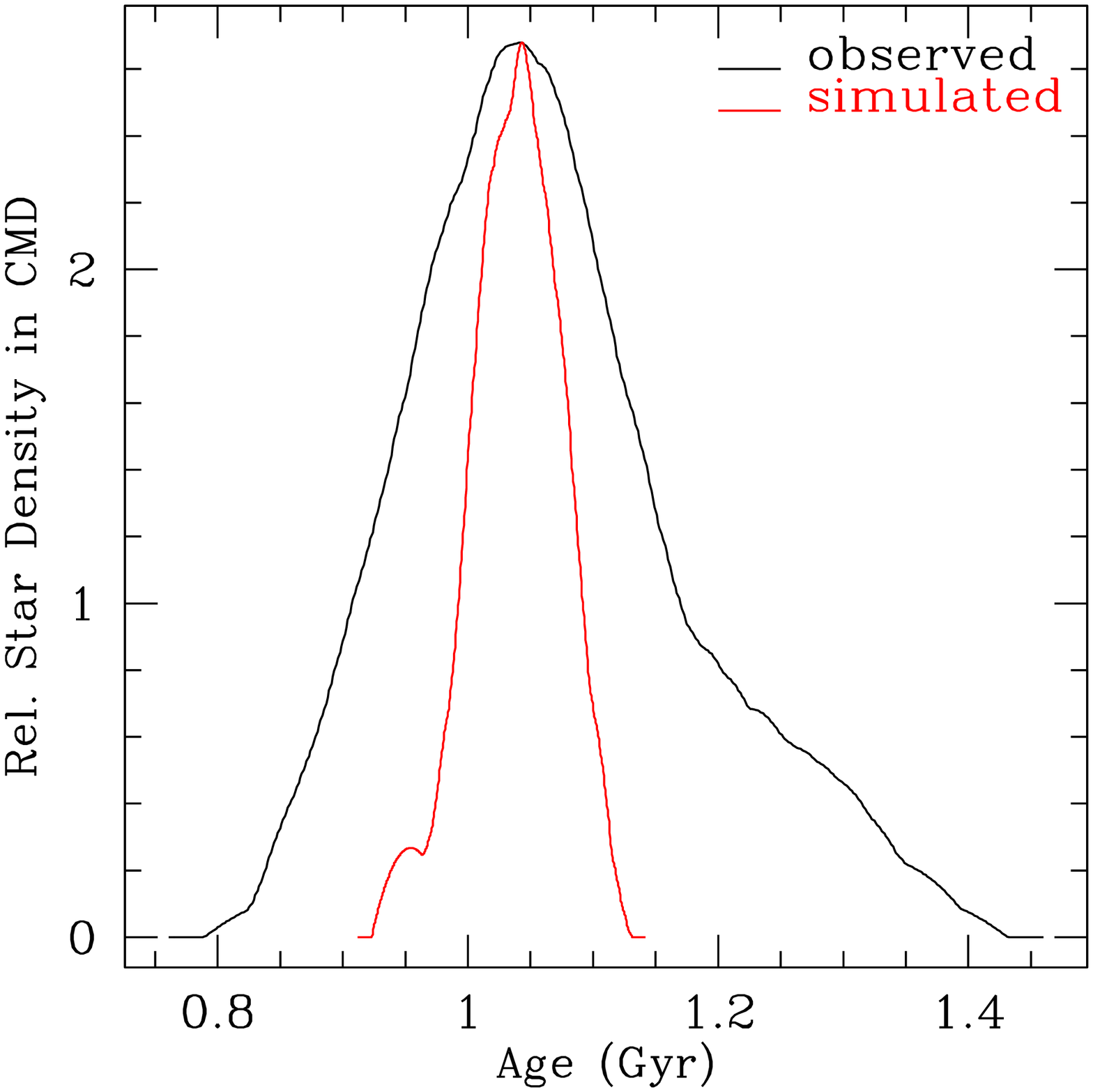}
\includegraphics[scale=0.29]{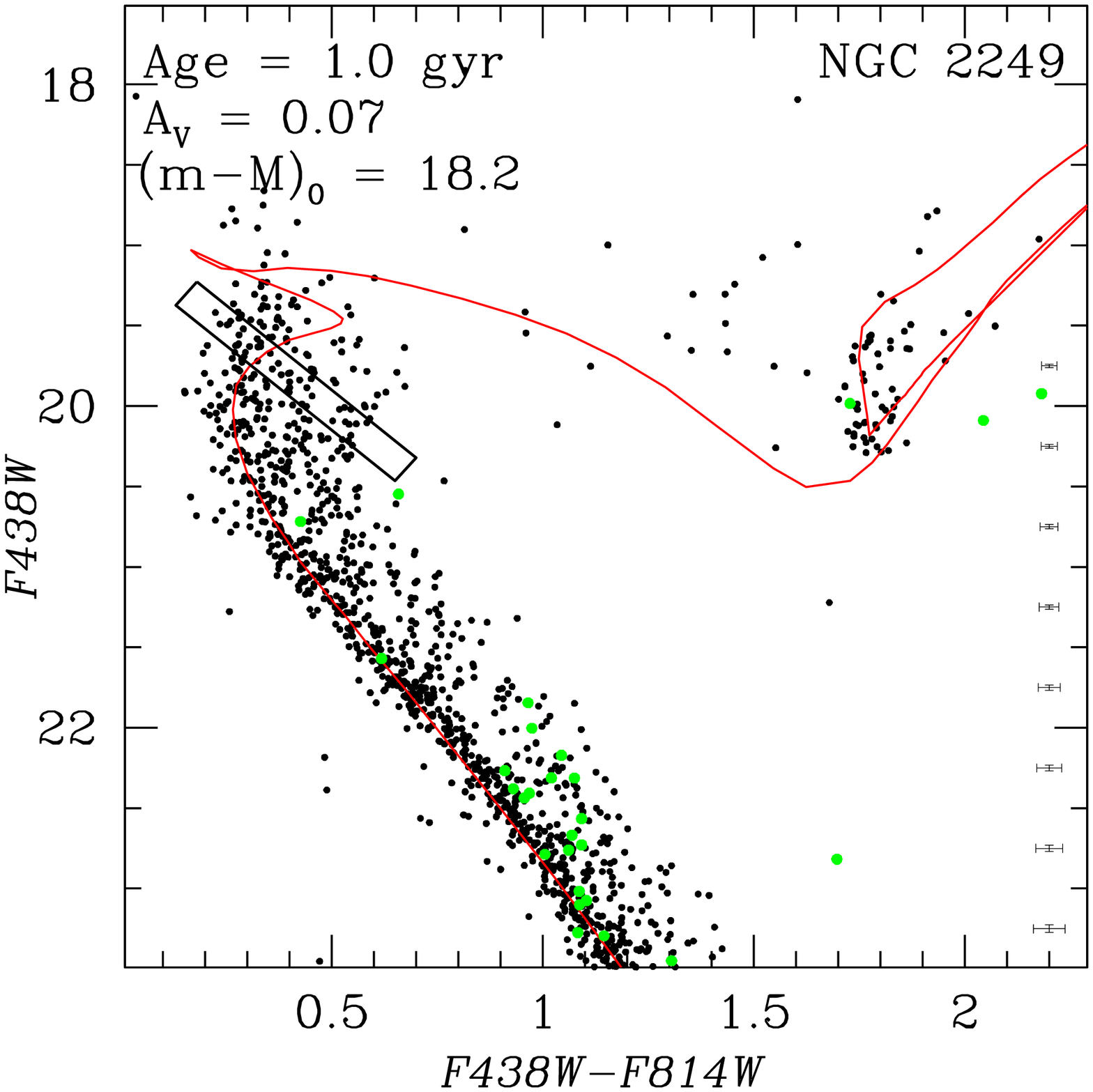}
\includegraphics[scale=0.29]{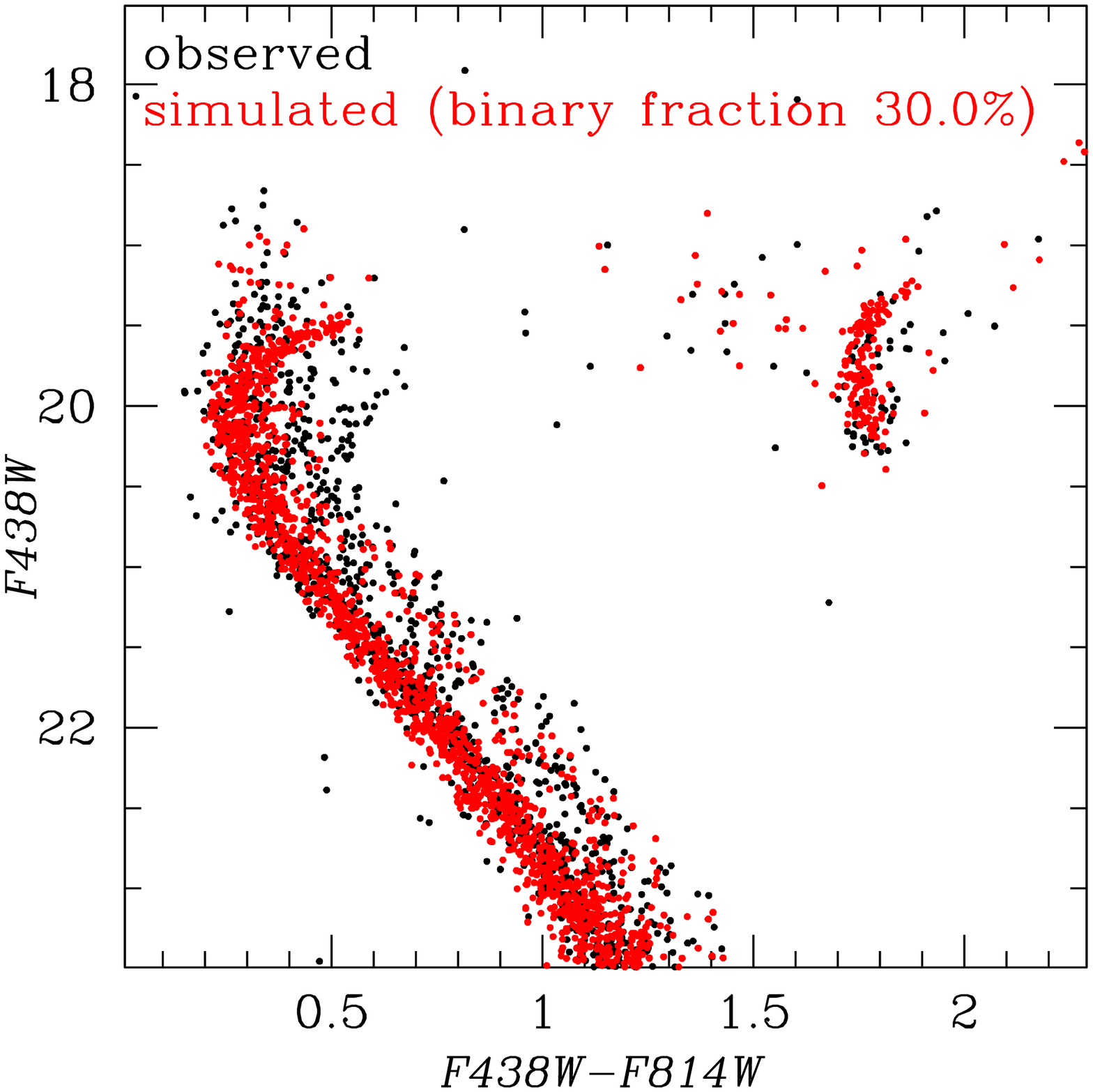}
\includegraphics[scale=0.29]{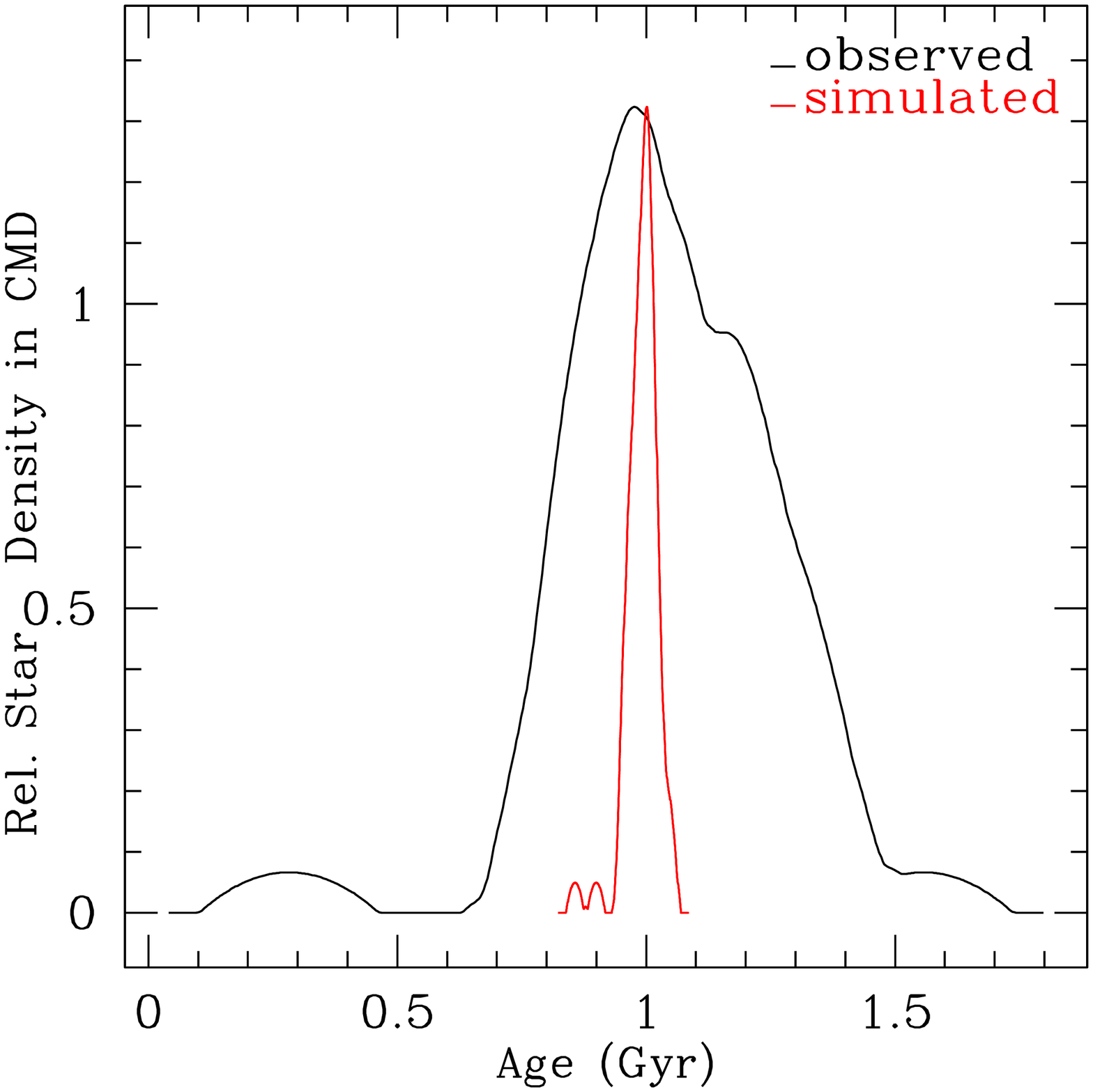}
\includegraphics[scale=0.29]{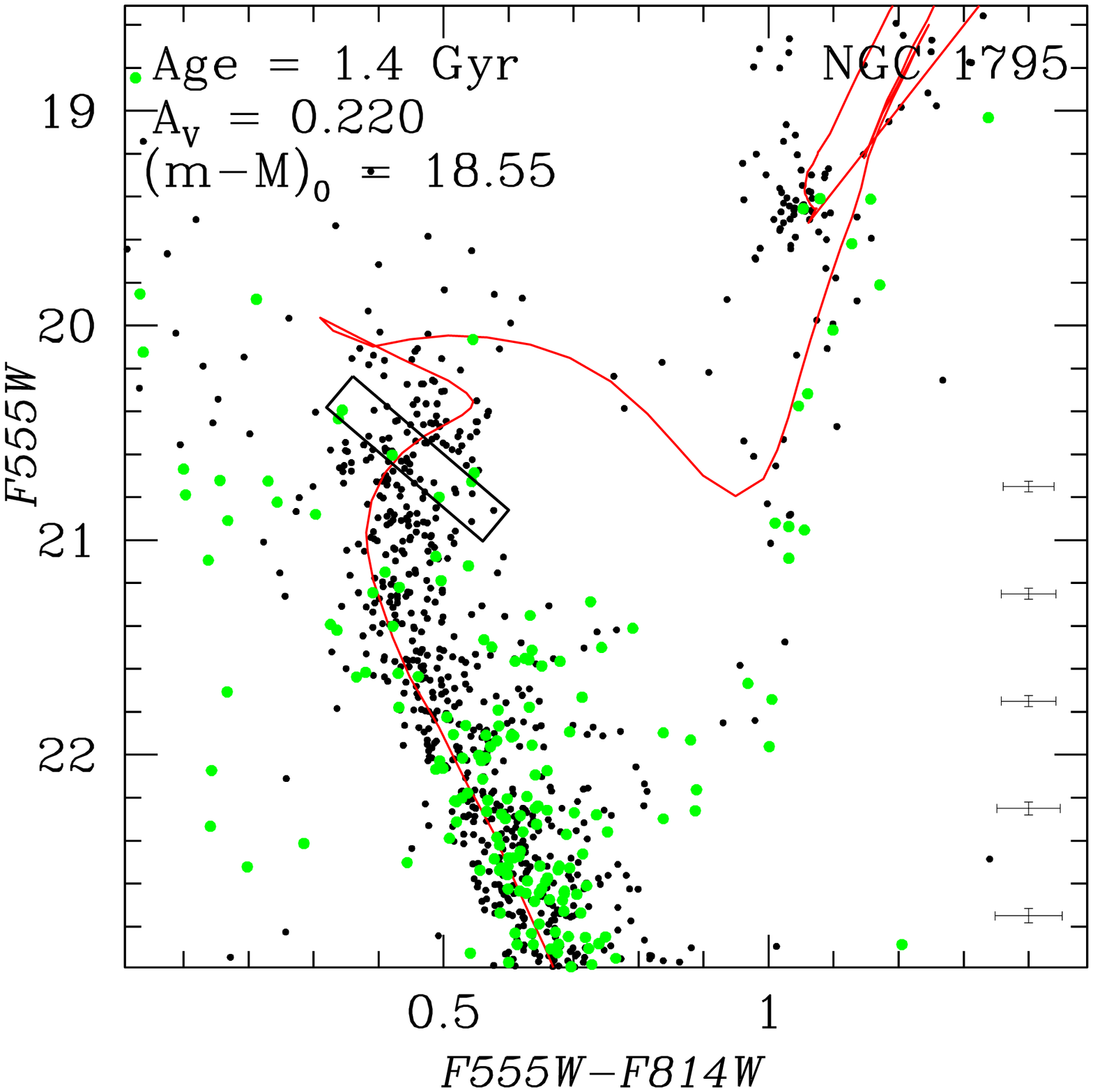}
\includegraphics[scale=0.29]{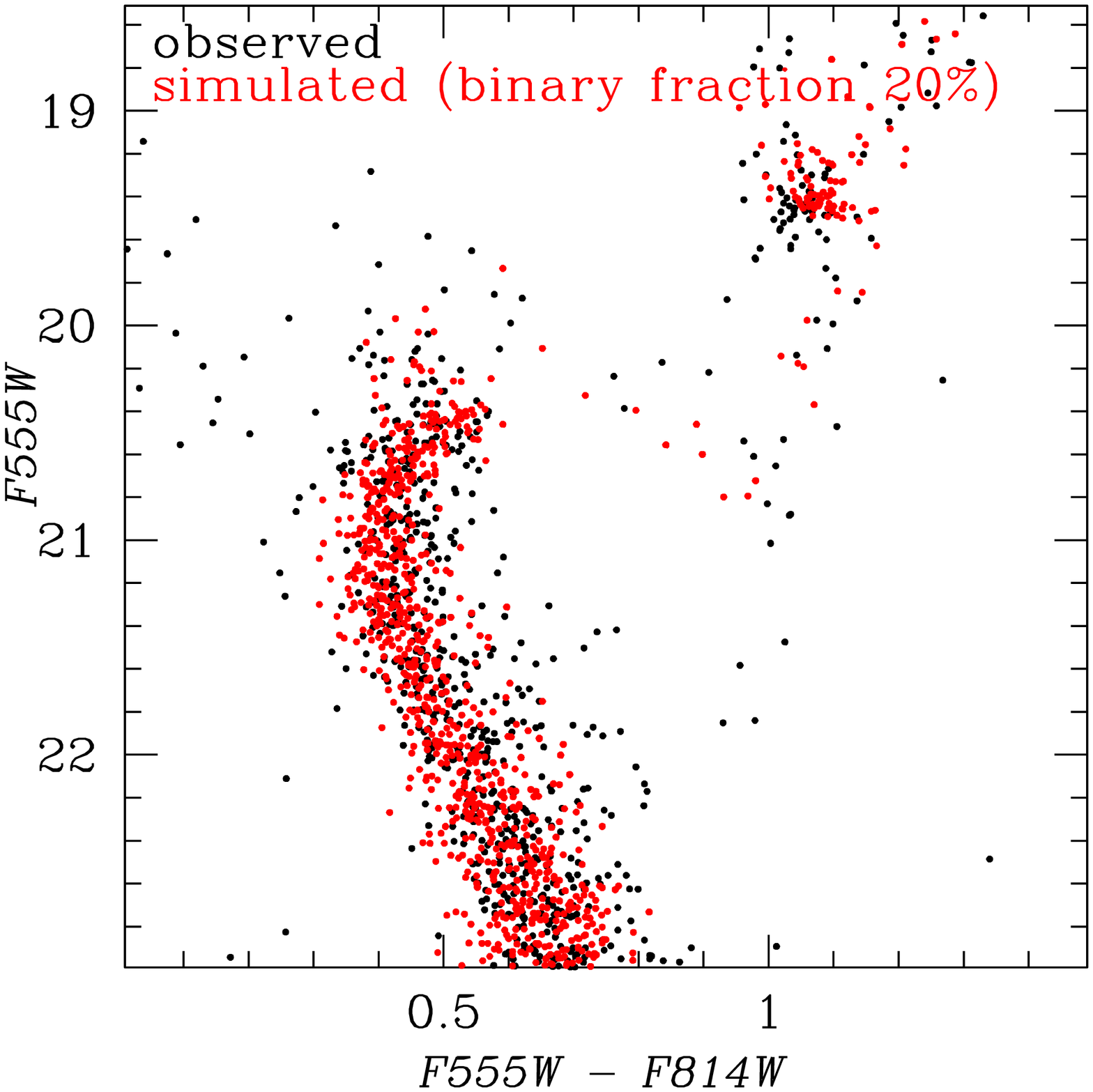}
\includegraphics[scale=0.29]{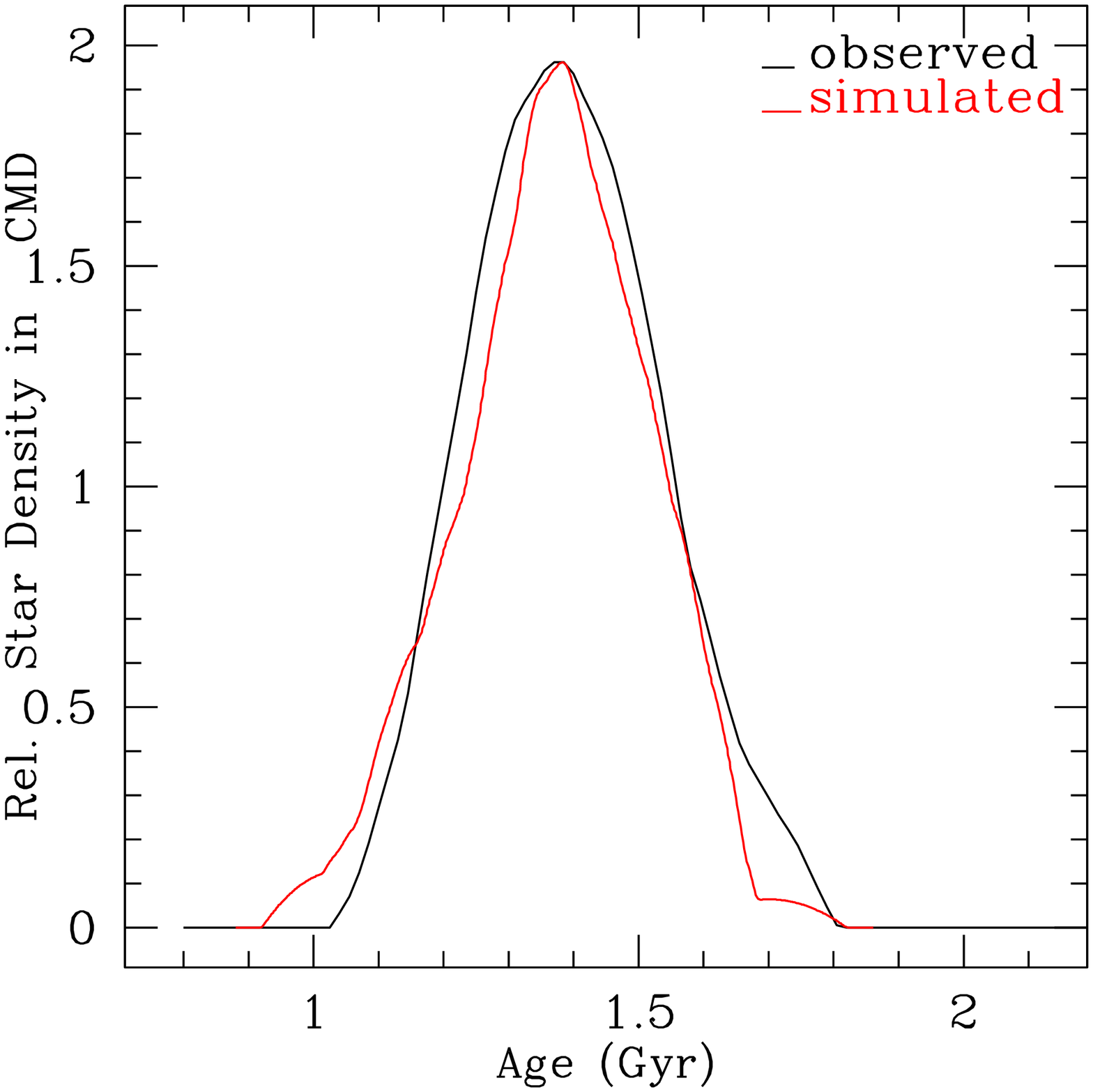}
\includegraphics[scale=0.29]{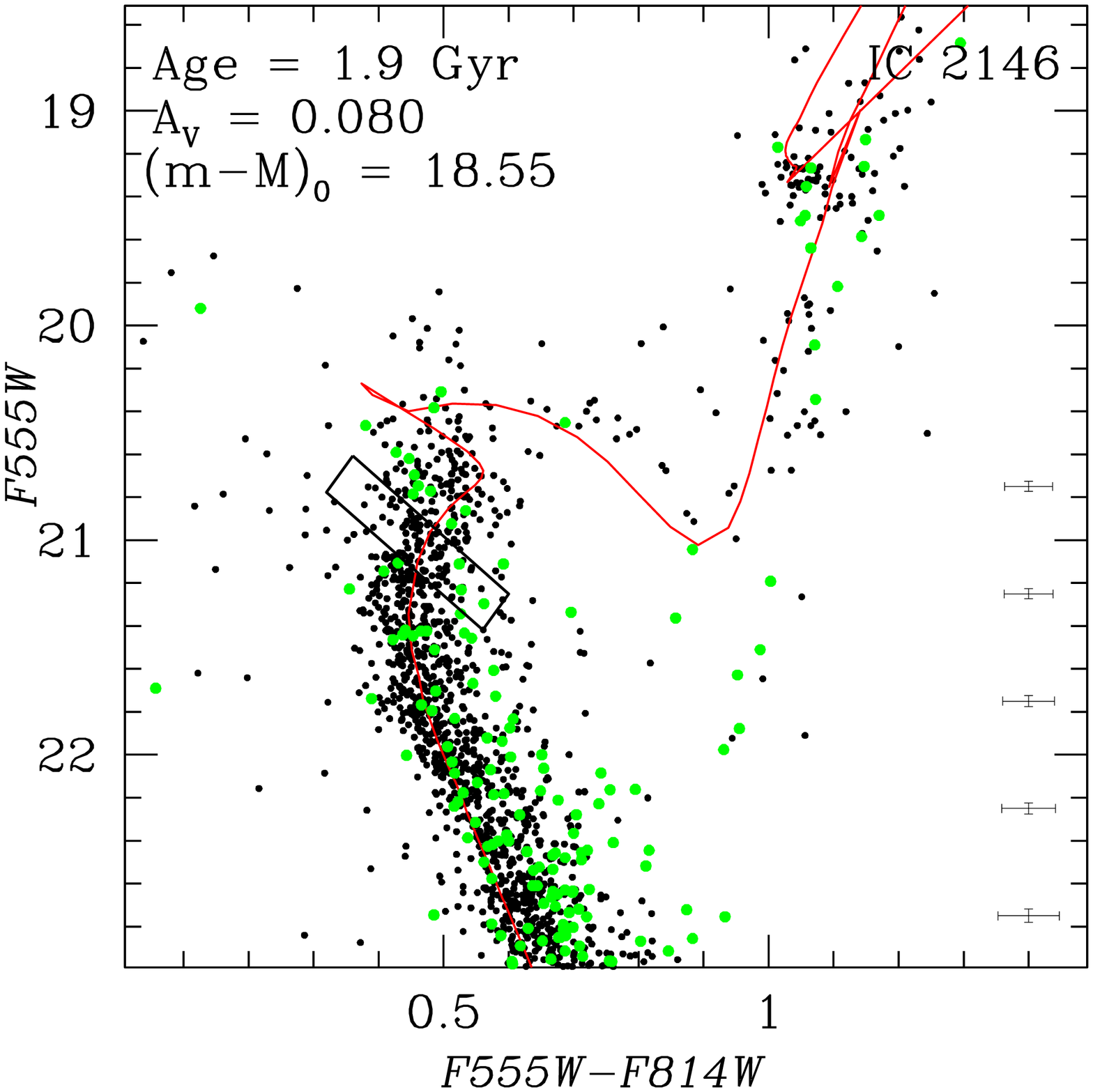}
\hspace{0.8mm}
\includegraphics[scale=0.29]{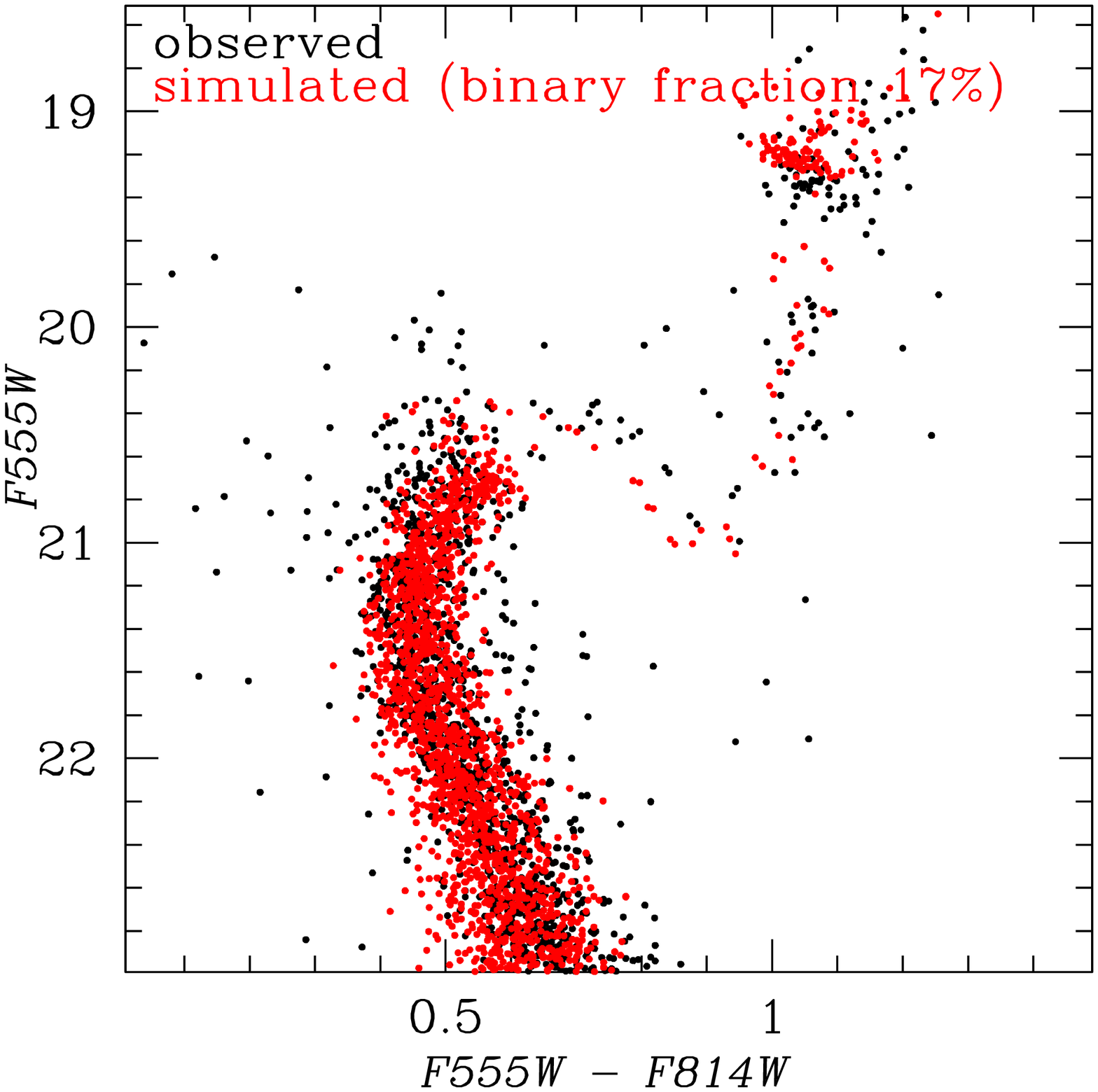}
\hspace{0.3mm}
\includegraphics[scale=0.29]{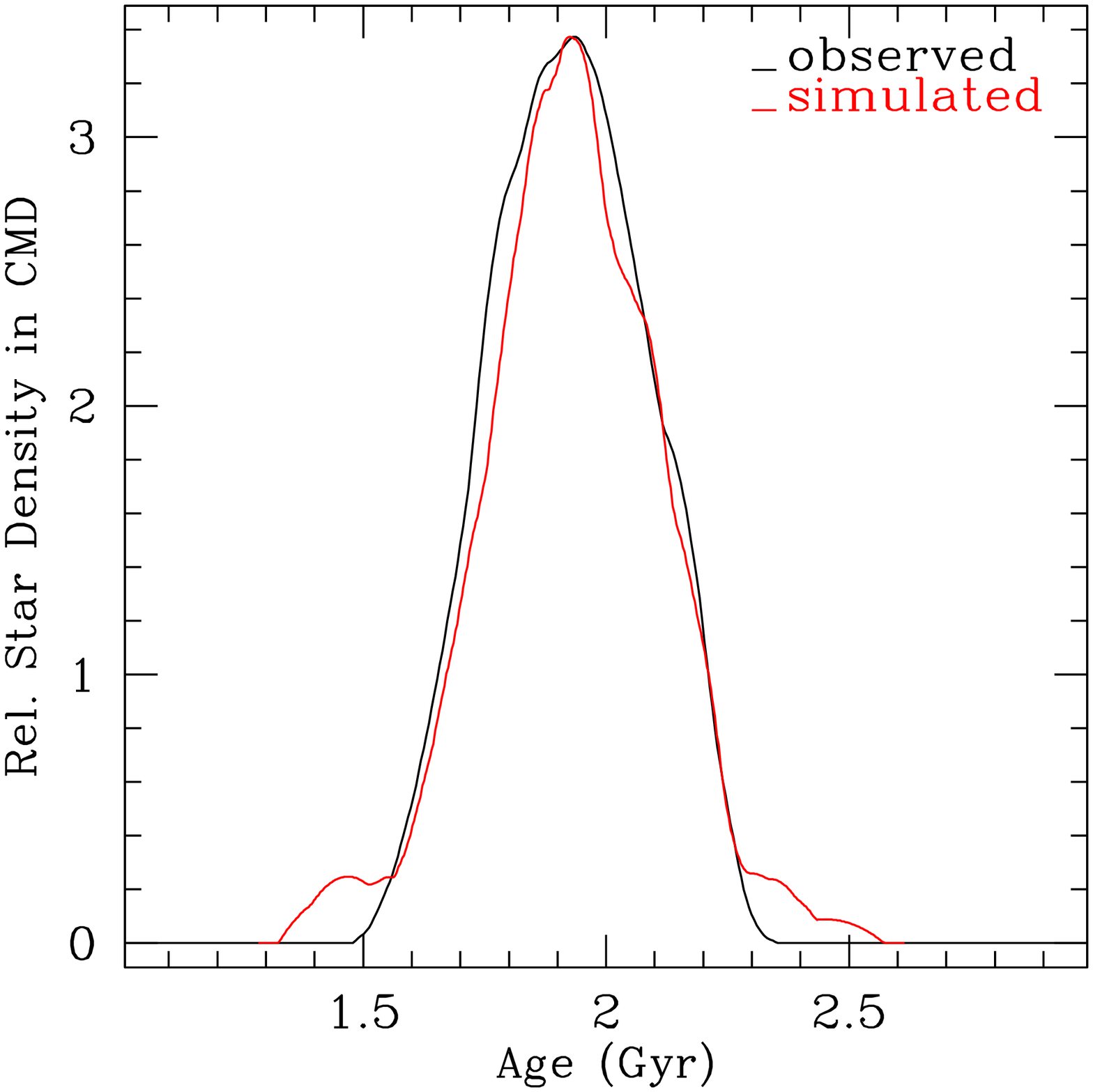}
\caption{Left panels: {\B} versus {\BI} or {\V} versus {\VI} CMDs for the four
star clusters. Best-fit isochrones from \cite{mari+08} are superposed to the
clusters' CMDs, along with the derived age, distance modulus $(m-M)_0$ and
visual extinction $A_V$. Contamination from the underlying LMC field population
has been derived from a region near the corner of the image, with the same
surface area adopted for the cluster stars, and superposed on the clusters' CMDs
(green dots). Parallelogram boxes used to select MSTOs stars to derive the
Pseudo-age distribution are also indicated. For each CMD we also show
magnitude and color errors, derived using the photometric distribution of our
artificial stars test. Center panels: comparison between observed (black dots)
and simulated (red dots) CMDs. For each simulated cluster we reported also the
adopted binary stars fraction. Right panels: Pseudo-age distributions, derived
as described in G11a, for the MSTO regions of the observed (black line) and
simulated (red line) CMDs.} 
\label{f:cmd}
\end{figure*}

CMDs for the four star clusters are presented in the left panels of
Fig.~\ref{f:cmd}. We plot the stars within a core radius (except for NGC~2249,
for which we plot stars within the effective radius $r_e$), based on the King
(1962) model fits (derived as described in Sect.~\ref{s:dynamics} below).
Best-fit isochrones from \cite{mari+08} are superposed onto the clusters' CMDs,
along with the derived age, distance modulus and visual extinction $A_V$. Isochrone fitting has been performed using the methods described in
\citep[][G11a]{goud+09}. Briefly, we used the observed difference in magnitude
between the MSTO and the red clump (RC); we selected all the isochrones for
which the value of this parameter lies within $2\sigma$ of the measured
uncertainty of that parameter in the CMDs. For the set of isochrones that
satisfy our selection criteria (on average 5\,--\,10 isochrones), we found the
best-fit values for distance modulus and reddening, by means of a least square
fitting program to the magnitudes and colors of the MSTO and RC. Finally, we
overplotted the isochrones onto the CMDs and through a visual examination we
selected the best-fitting one. To assess the level of contamination of the
underlying  LMC field population, we selected regions near the corners of the
image, with the same surface area as that adopted for the cluster stars. Stars
located in these regions have been superposed on the clusters' CMDs (shown as
green dots in the left panels of Fig~\ref{f:cmd}). The contamination in NGC~2209
and NGC~2249 is very low, while it is more pronounced in NGC~1795 and IC~2146,
but for all the four clusters is mainly confined to the lower (faint) part of
the clusters' MS.

Fig.~\ref{f:cmd} shows that the MSTO regions of NGC~2209 and NGC~2249 are fairly
wide, whereas the MS of single stars fainter than the turnoff is well defined
and much narrower. This difference is even more evident if we compare the MSTO
regions of these clusters with that of NGC~1795 and IC~2146. These two clusters
have a quite compact MSTO, for which the morphology seems consistent with that
of a single stellar population. Magnitude and color errors are shown in
the left panels of Fig.~\ref{f:cmd}: photometric errors at the MSTO magnitude
level are between 0.02 and 0.04 mag in color (F438W\,--\,F814W or
F555W\,--\,F814W, ), far too small to account for the broadening of the MSTO
region seen in NGC~2209 and NGC~2249. Moreover, reddening values are quite small
and the narrow morphology of the RC features shows that
differential reddening effects are negligible. Hence, the width of the MSTO in
these two clusters seems to strongly exclude a simple stellar population, and in
the context of the scenario described in the introduction, indicate the presence
of multiple stellar generations.
To constrain this picture further in terms of whether or not a single population
can reproduce the observed CMDs and to quantify the width of the eMSTO, we
conducted Monte Carlo simulations of synthetic clusters with the properties
implied by the isochrones fitting. 

We simulated a SSP with a given age and chemical composition by populating an
isochrone with stars randomly drawn from a Salpeter mass function and normalized
to the observed number of stars according to the completeness. To a fraction of
these sample stars, we added a component of unresolved binary stars drawn from
the same mass function, using a flat distribution of primary-to-secondary mass
ratios. To estimate the binary fraction in each cluster, we used the width of
the upper MS, i.e. the part between the MSTO region and the TO of the background
stellar population. We estimated that the internal systematic uncertainty
in the binary fraction is of the order of 5\%; for the purpose of this work, the
results do not change significantly within $\sim$ 10\% of the binary fraction.
Finally, we added photometric errors, derived using the photometric error
distribution of our artificial stars test. 

The comparison between the observed and simulated CMDs is shown in the middle
panels of Fig.~\ref{f:cmd}. The adopted binary star fraction is also reported
for each cluster. Overall, the SSP simulations reproduce several CMDs features
quite well in the four clusters, such as for example the MS and the RC. However,
the middle panels of Fig.~\ref{f:cmd} clearly show that the MSTO regions in
NGC~2209 and NGC~2249 are significantly wider than that of their respective SSP
simulations. Conversely, the MSTO regions of NGC~1795 and IC~2146 are very well
reproduced by their SSP simulation. 

In order to compare in detail the observed and simulated MSTO regions, we
created their ``pseudo-age'' distributions (see G11a for a detailed
description). Briefly, pseudo-age distributions are  derived by constructing
parallelograms across the MSTOs with one axis approximately parallel to the
isochrones and the other approximately perpendicular to them (parallelograms for
each clusters are illustrated in left panels of Fig.~\ref{f:cmd}). The ({\BI},
{\B}) and ({\VI}, {\V}) coordinates of the stars within the parallelogram are
then transformed into the reference coordinate frame defined by the two axes of
the parallelogram. The same procedure is then  applied to the isochrone tables
to set the age scale along this vector. A polynomial least-square fit between
age and the coordinate in the direction perpendicular to the isochrones then
yields the pseudo-age distributions. This procedure is applied both to the
observed and simulated CMDs. 

The derived pseudo-age distributions are shown in
the right panels of Fig.~\ref{f:cmd}. These were calculated using the
non-parametric Epanechnikov-kernel probability density function \citep{silv86},
in order to avoid biases that can arise if fixed bin widths are used. As
expected, for NGC~1795 and IC~2146 the observed pseudo-age distributions are
very well reproduced by the simulated ones. Conversely,  the observed pseudo-age
distributions for NGC~2209 and NGC~2249 are significantly wider than the
simulated ones: the stellar age distributions peak near the location of the
best-fit isochrone and then decline slower than the prediction of the single
stellar generation distributions. They do so slower towards older ages than
towards younger ages. The pseudo-age distributions of the observed clusters are
quite smooth,  suggesting that the morphologies of the MSTO regions can be
better explained with a spread in age rather than by two discrete  bursts of
star formation. This is similar to the findings of G11a for their 7 clusters.

Our detection of an eMSTO in NGC~2209 confirms the finding by \citet{kell+12}
who used $g$ and $i$-band imaging taken at Gemini-South. Due to the crowding in
the inner regions at ground-based spatial resolution, \citet{kell+12} only used
stars in an annulus with radii $40'' < r < 80''$ from the cluster center, which
is outside the cluster's half-light radius. Our results show that the eMSTO
feature in NGC~2209 extends all the way into the core, where the relative
contamination by field stars is much smaller. 

As to the non-detection of an eMSTO in NGC~1795 and IC~2146, it was shown by
\citet{kell+11} and G11a that the width of pseudo-age distributions  of
simulated SSPs scales approximately with the logarithm of the cluster age.  This
renders the ability to detect a given age spread to be age dependent, becoming
harder for older clusters. However, at the age of the oldest cluster in our
sample (IC~2146, 1.9 Gyr), an age spread $\Delta \tau$ = 200 Myr yields
$\Delta\tau/\tau$ = 0.105, which is well above the detection limit for the data 
(see G11a). This confirms the simple stellar population nature of NGC~1795 and
IC~2146. 

\section{Insights From Histories of Cluster Mass Loss and Escape Velocity} 
\label{s:dynamics}
As stated in the Introduction, the photometric evidence available to date led to
two main types of scenarios to explain the eMSTO phenomenon:\ scenarios that
involve the formation of a second generation of stars, and scenarios that do
not. The former type of scenarios consists of two subtypes: \emph{(i)} the
scenario of K11 who suggested that eMSTO can only be hosted by intermediate-age
star clusters with current core radii larger than a certain threshold value
(i.e.. $r_c \ga$ 3.7 pc). We refer to this scenario as the ``core radius
threshold'' scenario hereinafter; and \emph{(ii)} the ``escape velocity
threshold'' scenario of G11b who proposed that eMSTO can only be hosted by
clusters for which the escape velocity of the cluster was higher than the wind
velocities of the stars thought to provide the material used by the second
generations of stars, at the time such stars were present in the cluster. The
scenarios that do not involve extended star formation consist of \emph{(i)} the
``stellar rotation'' scenario first suggested by \citet{basdem09} and
\emph{(ii)} the ``interactive binaries'' scenario of \citet{yang+11}. Noting
that these different scenarios imply different predictions on the dependence of
eMSTO properties on the clusters' dynamical properties, we proceed to determine
structural parameters and masses of our sample clusters. 

\subsection{Structural Parameters of the Clusters}
\label{s:kingmodels}
\begin{figure*}[thp]
\centerline{
\includegraphics[height=8.2cm,angle=270]{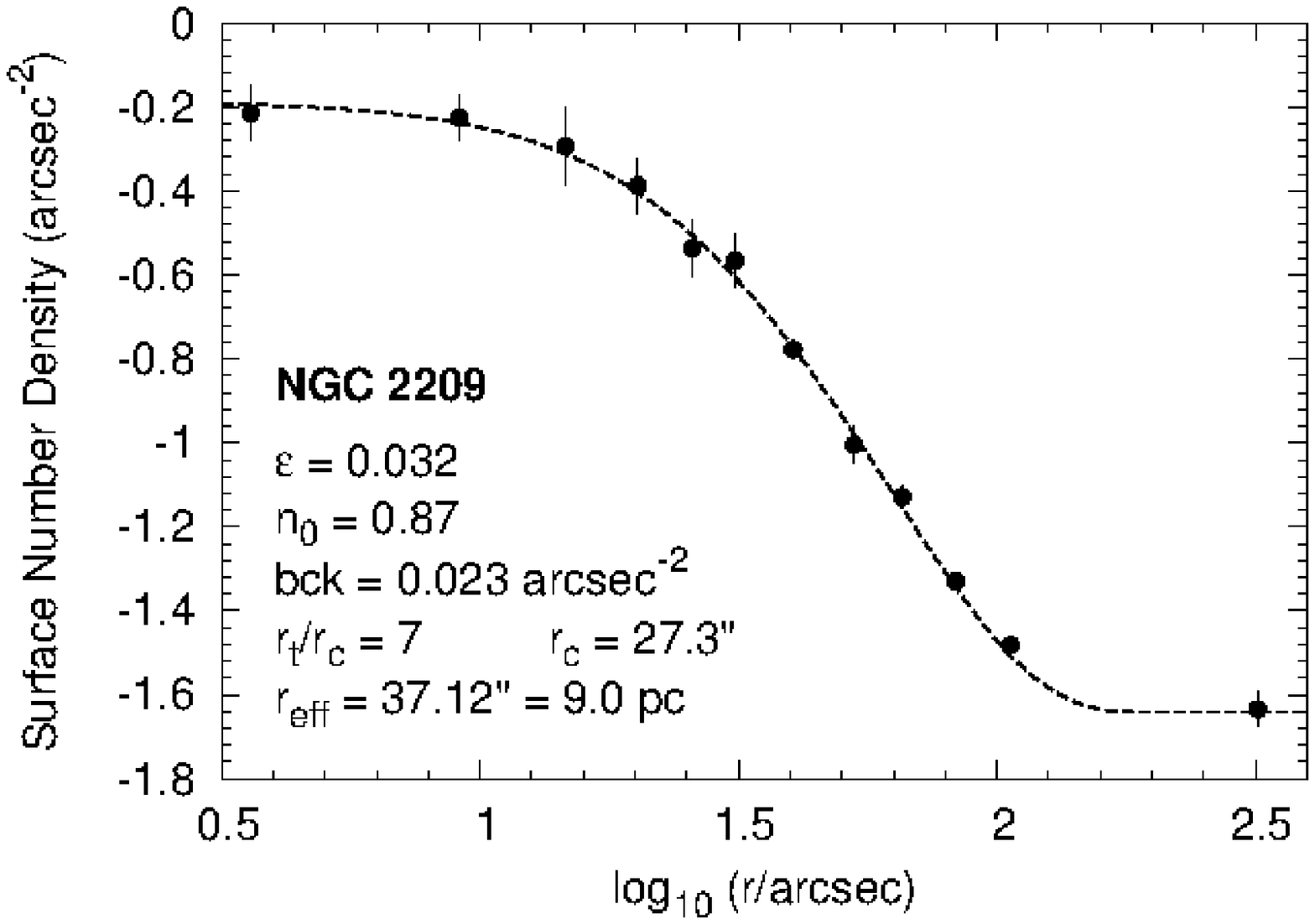}
\hspace*{1mm}
\includegraphics[height=8.2cm,angle=270]{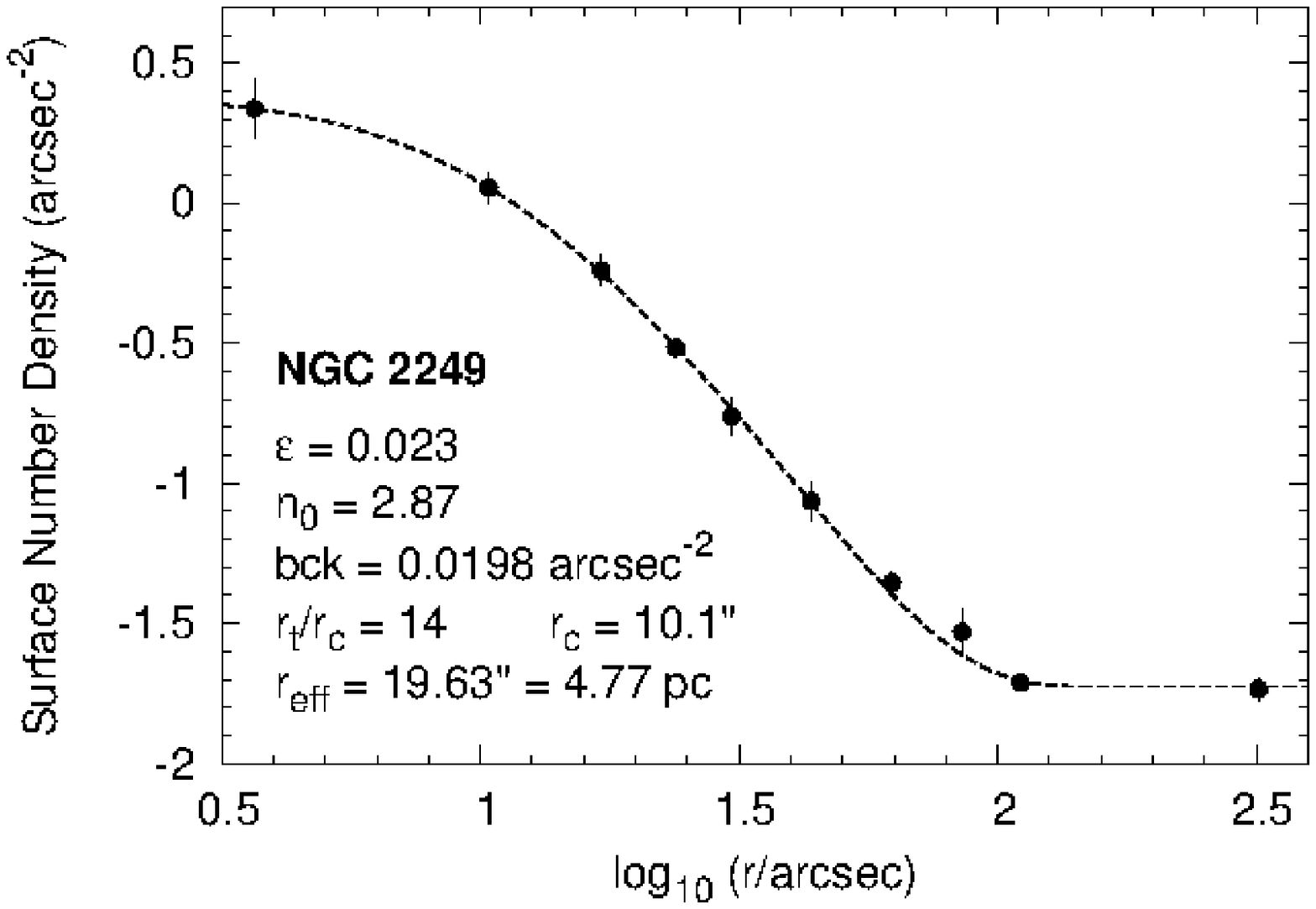}
}
\centerline{
\includegraphics[height=8.2cm,angle=270]{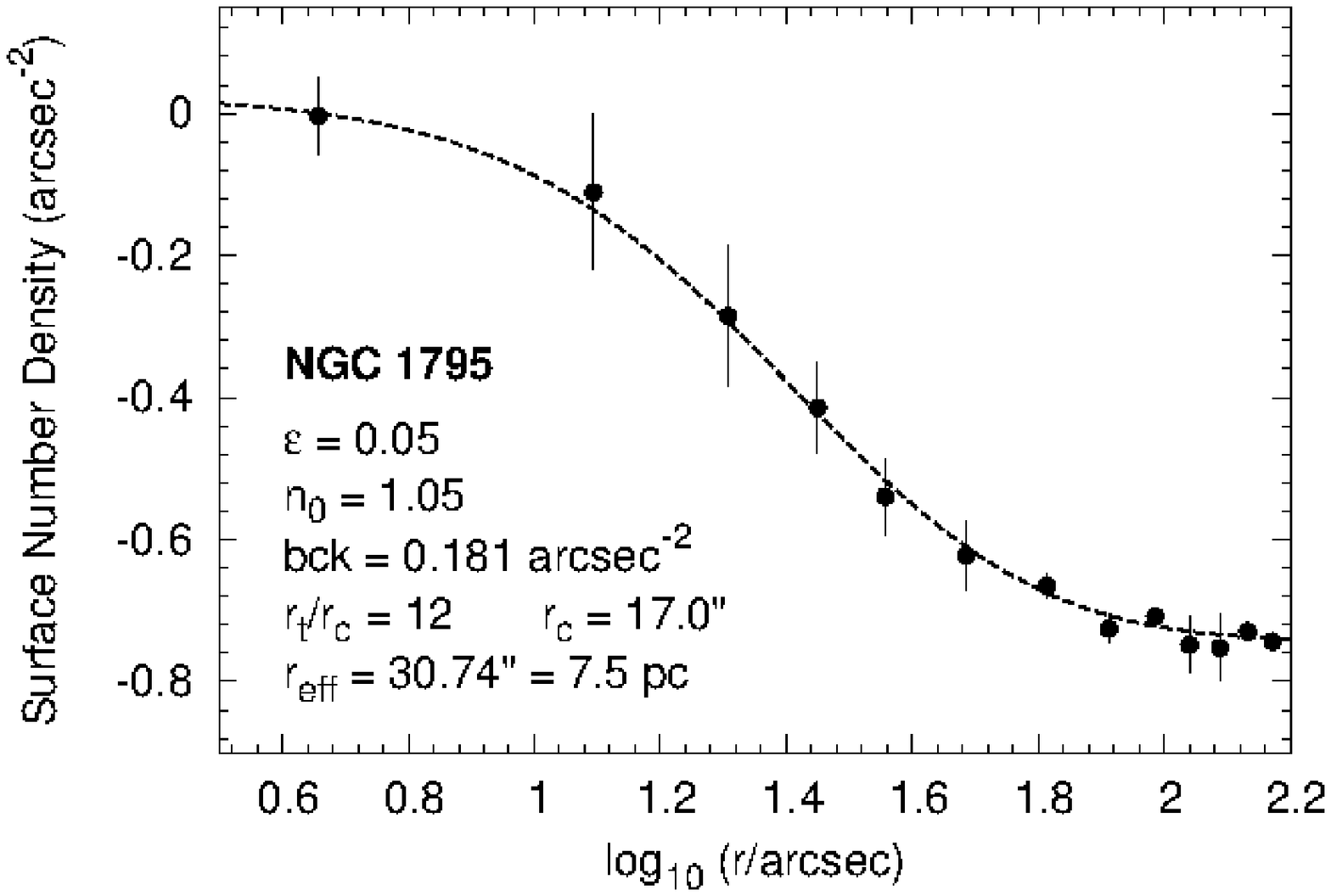}
\hspace*{1.0mm}
\includegraphics[height=8.2cm,angle=270]{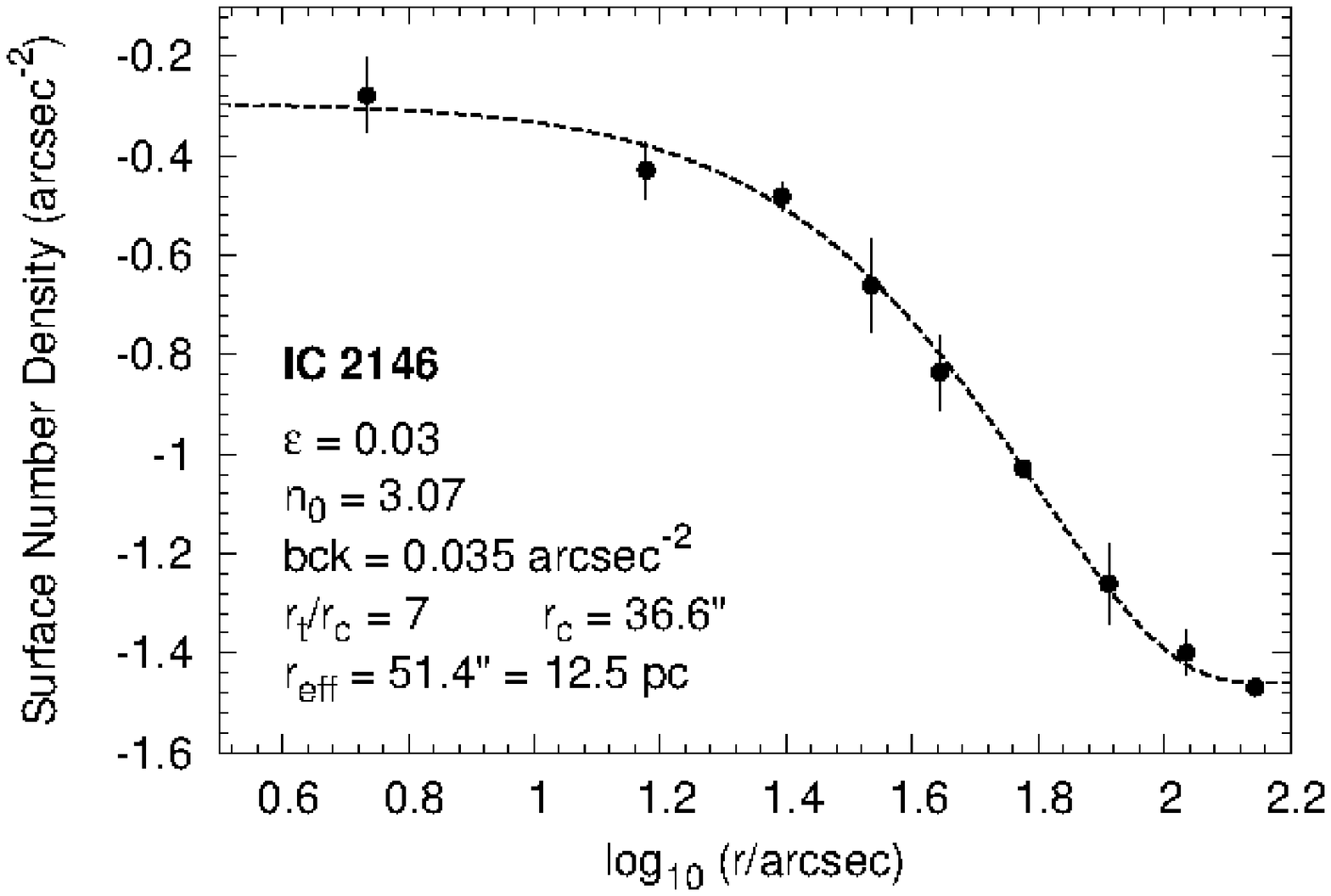}
}
\caption{Radial surface number density profiles of the four star clusters in our
sample. The points represent observed values. The dashed lines represent the
best-fit King models (cf. Equation~\ref{eq:King}) whose parameters are shown in
the legend. Names, ellipticities and effective radii of the clusters are also
shown in the legend. The radius values have been converted to parsec adopting the appropriate distance modulus.} 
\label{f:king}
\end{figure*}

We first determine completeness-corrected radial number density distributions of
stars for the four clusters in our sample following \citet{goud+09}. We
determine the clusters' center creating a two dimensional histogram of the pixel
coordinates using a bin size of $50 \times 50$ pixels, and then calculating the
center using a two-dimensional Gaussian fit to an image constructed from the
surface number density values in the two-dimensional histogram. This method
avoids biases related to the presence of bright stars near the center. The 
cluster's ellipticity $\epsilon$  is derived by running the task {\it ellipse}
within IRAF/STSDAS\footnote{STSDAS is a product of the Space Telescope Science
Institute, which is operated by AURA for NASA.} on the surface number density
images mentioned above. The area sampled by the images is then divided in a
series of concentric elliptical annuli, centered on the clusters center. The
spatial completeness of each annulus is divided out in the process. For NGC~2209
and NGC~2249 the outermost data point is derived  from the ACS parallel
observations, in a field located $\simeq$\,5\farcm5 from the cluster center. We
fit the radial surface number density profile using a King (1962) model combined
with a constant background level, described by the following equation: 
\begin{equation}
n(r) = n_0 \: \left( \frac{1}{\sqrt{1 + (r/r_c)^2}} - \frac{1}{\sqrt{1+c^2}}
\right)^2 \; + \; {\rm bkg} 
\label{eq:King}
\end{equation}
where $n_0$ is the central surface number density, $r_c$ is the core radius,  $c
\equiv r_t/r_c$ is the King concentration index ($r_t$ being the tidal radius),
and $r$ is the geometric mean radius of the ellipse ($r = a\,\sqrt{1-\epsilon}$,
where a is the semi-major axis of the ellipse). The best-fit King models for
the four clusters, selected using a $\chi^2$ minimization routine, are shown in
Fig.~\ref{f:king}, along with the derived surface number density values and
other relevant  parameters. 
\begin{figure}[tbh]
\includegraphics[width=1\columnwidth]{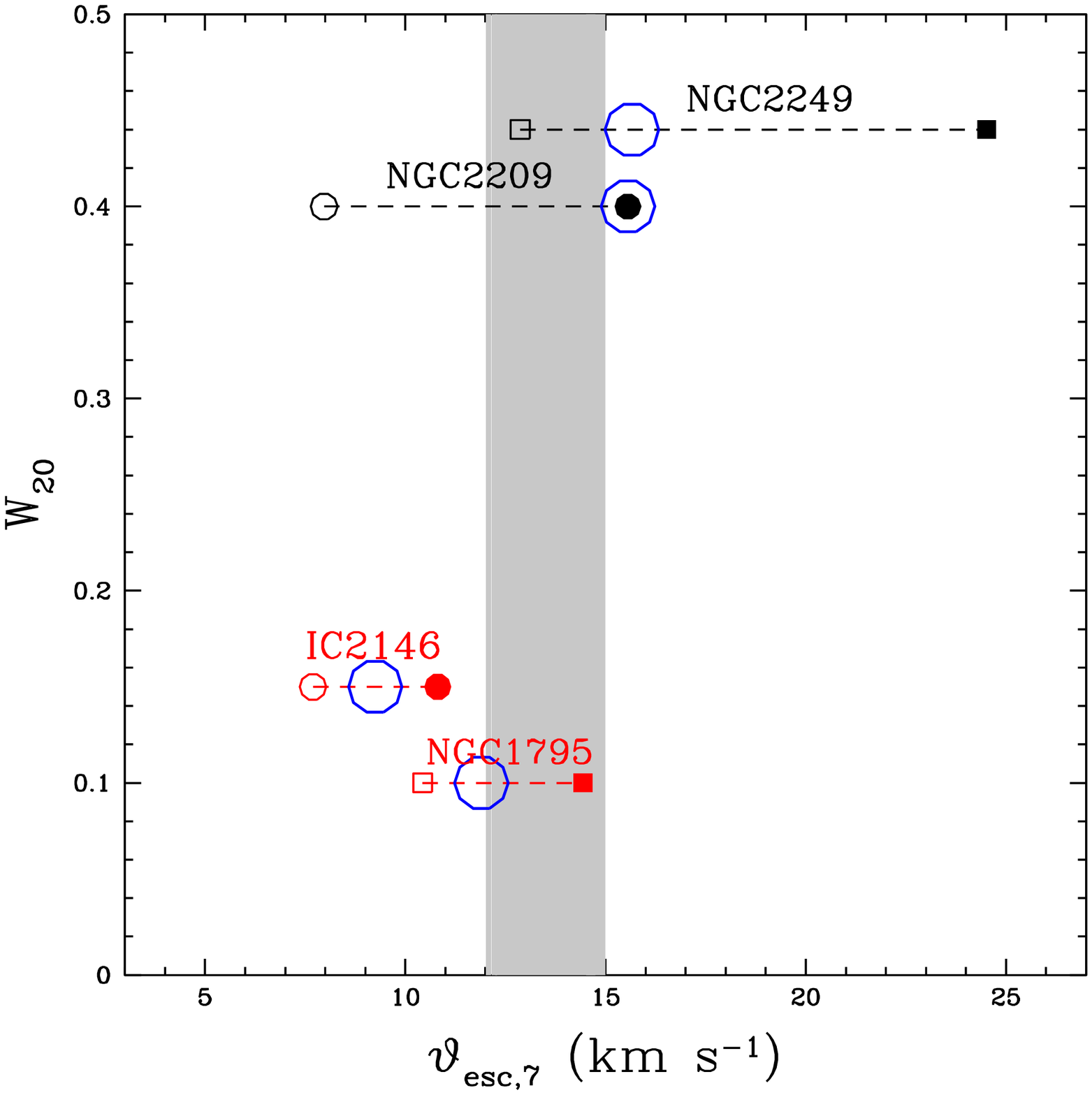}
\caption{The parameter $W_{20}$, that is the width at the 20\% of the maximum
value of the clusters' pseudo age distributions, after deconvolving out the
values for the SSP simulations and represents a measure of the broadening of the
MSTO, is plotted against the escape velocity at an age of $10^7$ Myr (the
derived $W_{20}$ value for NGC~1795 and IC~2147 is an upper limit). Black (red)
points represent clusters exhibiting (not exhibiting) an eMSTO; currently
extended (compact) clusters are shown with circles (squares). Finally, open
(filled) symbols represent data using a model without (with) initial mass
segregation. The blue open circles represent the most ``plausible'' escape
velocities values. The grey region depicts the range of escape velocities within
which the distinction between clusters with and without eMSTOs seems to occur
(see discussion in Section~\ref{s:dynevol}).}  
\label{f:escvel}
\end{figure}
Fig.~\ref{f:king} shows some interesting results. First of all, the two clusters
hosting an eMSTO, NGC~2209 and NGC~2249, have a very different core radius.
While NGC~2209 is currently extended (i.e.: $r_c = 6.64$ pc), NGC~2249 is
currently compact (i.e.: $r_c = 2.46$ pc). Second, the core radii of the two
clusters in our sample that do not exhibit an eMSTO, NGC~1795 and IC~2146, are
also interesting in the context of constraining different scenarios. While
NGC~1795 has a moderately large core radius ($r_c = 4.13$ pc), IC~2146 is
\emph{very} extended ($r_c = 8.89$ pc, the most extended in our sample).  Taking
these results at face value, it seems that  the ``core radius threshold''
scenario can not fully account for the presence of eMSTOs in intermediate-age
star clusters. Specifically, under the hyphothesis that a cluster can host an
eMSTO only if it has a radius larger than a certain value ($r_c \ga$ 3.7 pc), we
would have expected to detect eMSTOs in NGC~1795, NGC~2209, and IC~2146, and
\emph{not} in NGC~2249 for which the derived core radius is significantly
smaller than the threshold value. Instead, eMSTOs are \emph{not} detected in
NGC~1795 and IC~2146, while an eMSTO \emph{is} detected in NGC~2249.

\subsection{Present-Day Masses and Dynamical Evolution}
\label{s:dynevol}
To verify whether the ``escape velocity threshold'' scenario \emph{can} provide
a valid explanation on the formation of eMSTO, we estimate cluster masses and
escape velocities as a function of time, going back to an age of 10 Myr, after
the cluster has survived the era of violent relaxation and when the most massive
stars of the first generation, proposed to be candidate polluters in the
literature (i.e., FRMS and massive binary stars), are expected to start losing
significant amounts of mass through slow winds.

Current cluster masses are determined from integrated-light $V$-band magnitudes
listed in Table~\ref{t:parameters}. Aperture corrections for these magnitudes
are determined from the best-fit King model for each cluster by calculating the
fraction of total cluster light encompassed by the measurement aperture. After
aperture correction, cluster masses are calculated from the values of $A_V,
(m\!-\!M)_0, \mbox{[Z/H]}$, and age listed in Table~\ref{t:parameters}. This is
done by interpolation between the ${\cal{M}}/L_V$ values in the SSP model tables
of \citet{bc03}, assuming a \citet{salp55} initial mass function.
\begin{table*}[thb]
\begin{center}
\caption{Physical properties of the star clusters}
\begin{tabular}{ccccccccc}
\hline
\hline
Cluster & V & Aper. & Aper. corr. & [Z/H] & $A_V$ & Age & $r_c$ & $r_e$ \\
 (1) & (2) & (3) & (4) & (5) & (6) & (7) &  (8) & (9) \\
\hline
NGC~2209 & 13.15 $\pm$ 0.15 & 31 & 1.00 & -0.30 & 0.23 & 1.15 & 6.64 $\pm$ 0.34 & 9.02 $\pm$ 0.46 \\
NGC~2249 & 12.23 $\pm$ 0.15 & 31 & 0.26 & -0.46 & 0.07 & 1.00 & 2.46 $\pm$ 0.04 & 4.77 $\pm$ 0.08 \\
NGC~1795 & 12.67 $\pm$ 0.15 & 31 & 0.63 & -0.30 & 0.20 & 1.40 & 4.13 $\pm$ 0.61 & 7.47 $\pm$ 1.23 \\
IC~2146  & 13.10 $\pm$ 0.15 & 50 & 0.76 & -0.30 & 0.08 & 1.90 & 8.89 $\pm$ 1.36 & 12.53 $\pm$ 1.92 \\
\hline
\end{tabular}
\tablecomments{Columns (1): Name of the clusters. (2): Integrated $V$ magnitude. (3): Adopted radius in arcsec for the measure of the integrated magnitude. (4): Aperture correction in magnitude. (5): Metallicity (dex). (6): Visual extinction $A_V$ in magnitude. (7): Age in Gyr. (8): Core radius $r_c$ in pc. (9): Effective radius $r_e$ in pc.}
\label{t:parameters}
\end{center}
\end{table*} 
Dynamical evolution calculations of the star clusters is done following the
prescriptions of G11b. Briefly, we evaluated the evolution of cluster mass and
effective radius for model clusters with and without initial mass segregation.
The latter property plays a fundamental role in terms of cluster mass loss and
the early evolution of cluster core radii \citep[e.g.,][]{mack+08b,vesp+09}.  
The dynamical evolution calculations cover an age range between 10 Myr and 13
Gyr and take into account the effects of stellar evolution mass loss and
internal two-body relaxation.  For each cluster, we listed in
Table~\ref{t:dynamics} the results obtained with a level of initial mass
segregation of $r_e / r_{e,1} = 1.5$ (where $r_{e,1}$ is the effective radius of
stars with $M > 1$ {\Msun}) versus those without initial mass segregation. 
Escape velocities listed in Table~\ref{t:dynamics} are determined for every
cluster by assuming a single-mass King model with a radius-independent
${\cal{M}}/L$ ratio\footnote{We acknowledge that this will underestimate
somewhat the $V_{\rm esc}$ values for clusters with significant mass
segregation.}   as calculated above from the clusters' best-fit age and [$Z$/H]
values. Escape velocities are calculated from the reduced gravitational
potential,  $V_{\rm esc} (r,t) = (2\Phi_{\rm tid} (t) - 2\Phi (r,t))^{1/2}$, at
the core radius. Here $\Phi_{\rm tid}$ is the potential at the tidal
(truncation) radius of the cluster. For convenience, we define $V_{\rm esc, 7}
(r) \equiv V_{\rm esc}\,(r, t = 10^7 {\rm yr})$, and refer to it as ``early escape  velocity''. 
\begin{table*}
\begin{center}
\caption{Dynamical properties of the star clusters}
\begin{tabular}{c|ccc|cccc}
\hline
 \colhead{} & \multicolumn{3}{|c|}{log (${\cal{M}}_{\rm cl}/M_{\odot}$)} & 
 \multicolumn{4}{|c}{$\vartheta_{\rm esc} \ $(\kms)}\\
Cluster & Current & 10$^7$ yr w/o m.s. & 10$^7$ yr with m.s. &  Current & 
 10$^7$ yr w/o m.s. & 10$^7$ yr with m.s. &  10$^7$ yr ``plausible''\\
(1) & (2) & (3) & (4) & (5) & (6) & (7) & (8)\\
\hline
NGC~2209 & 4.40 $\pm$ 0.07 & 4.53 $\pm$ 0.07 & 4.94 $\pm$ 0.07 & 6.1 $\pm$ 0.5 & 7.97 $\pm$ 0.5 & 15.56 $\pm$ 0.5 & 15.56 $\pm$ 0.5 \\
NGC~2249 & 4.48 $\pm$ 0.07 & 4.64 $\pm$ 0.07 & 5.04 $\pm$ 0.07 & 9.4 $\pm$ 0.8 &
12.87 $\pm$ 0.8 & 24.52 $\pm$ 0.8 & 15.66 $\pm$ 0.8 \\
NGC~1795 & 4.50 $\pm$ 0.07 & 4.66 $\pm$ 0.07 & 4.84 $\pm$ 0.07 & 7.7 $\pm$ 0.9 & 10.43 $\pm$ 0.9 & 14.44 $\pm$ 0.9 & 11.90 $\pm$ 0.9 \\
IC~2146 &  4.49 $\pm$ 0.07 & 4.63 $\pm$ 0.07 & 4.82 $\pm$ 0.07 & 5.9 $\pm$ 0.7 &
7.69 $\pm$ 0.7 & 10.81 $\pm$ 0.7 & 9.25 $\pm$ 0.7 \\
\hline
\end{tabular}
\tablecomments{Columns (1): Name of the clusters. (2): Logarithm of the adopted current cluster mass. (3-4): Logarithm of the adopted cluster mass at an age of 10$^7$ yr without(with) the inclusion of initial mass segregation. (5): Current cluster escape velocity at the core radius. (6-7): Cluster escape velocity at the core radius at an age of 10$^7$ yr without(with) the inclusion of initial mass segregation. (9) ``Plausible'' cluster escape velocity at an age of 10$^7$ yr.}
\label{t:dynamics}
\end{center}
\end{table*}
Early escape velocities for the four clusters, at an age of 10 Myr, are plotted
versus the $W_{20}$ parameter in Fig.~\ref{f:escvel}. $W_{20}$ represents the
width at the 20\% level with respect to the maximum value of the clusters pseudo
age distributions, after deconvolving out the values for the SSP simulations.
Currently extended and currently compact clusters are shown with different
symbols in Fig.~\ref{f:escvel}:\ the former as circle and the latter as square.
Furthermore, different colors are used to identify clusters hosting or not
hosting an eMSTO (black versus red, respectively). Finally, for each clusters,
we show the values for $V_{\rm esc,\,7}$ calculated without and with the
inclusion of initial mass segregation in the dynamical evolution model (open
versus filled symbols, respectively). To estimate ``plausible'' values for
$V_{\rm esc,\,7}$ for the different clusters we used a procedure that take into account the various results from the compilation of Magellanic Cloud star cluster properties and N-body simulations by \citet{mack+08b}. Briefly, they showed that the maximum core radius seen among a large sample of Magellanic Cloud star clusters increase approximately linearly with log(age) up to an age of $\sim$ 1.5 Gyr, namely from $\simeq$ 2.0 pc at $\simeq$ 10 Myr to $\simeq$ 5.5 pc at $\simeq$ 1.5 Gyr. Conversely, the {\it minimum} core radius is $\sim$ 1.5 pc throughout the age range 10 Myr\,--\,2 Gyr. Using N-body modeling \citet{mack+08b} showed that this behavior is consistent with adiabatic expansion of the cluster core in clusters with different levels of initial mass segregation, that is the cluster with the highest level of mass segregation experience the strongest core expansion. Finally, in our estimate we also considered the impact of the retention of stellar black holes (BHs) to the evolution of the clusters core radii (for a detailed description of the procedure, we refer the reader to a companion paper, Goudfrooij et al. 2014, submitted to ApJ).
Fig.~\ref{f:escvel} suggests that the differences among the clusters
can be explained by dynamical evolution arguments if the currently extended
clusters experienced stronger initial mass segregation than the currently
compact ones (to guide the reader, we highlighted with a blue circle the most
``plausible'' escape velocities values to be considered in the context of the
assumption mentioned above). 
\subsection{Implications on the nature of the eMSTOs}
\label{s:causes}
Under the assumption mentioned above, NGC~2209 and NGC~2249 have an early escape
velocity  $V_{\rm  esc} >$ 15 \kms, whereas the ones  derived for NGC~1795 and
IC~2146 are at $\sim$\,9 and 12 \kms. Taking these results at face value, the
``critical'' escape velocity above which the cluster is able to retain material
ejected by the slow winds of first  generation polluters seems to be in the
approximate range of 12\,--\,15 \kms\ (i.e., the grey area depicted in
Fig.~\ref{f:escvel}). This is in agreement with the results obtained by
G11b\footnote{Note however that G11b calculated their velocities using the
effective radius $r_e$ instead of the core radius $r_c$,  preventing a direct
comparison.} and Goudfrooij et al. (2014).

\begin{figure}[tbh]
\centerline{\includegraphics[width=0.7\columnwidth]{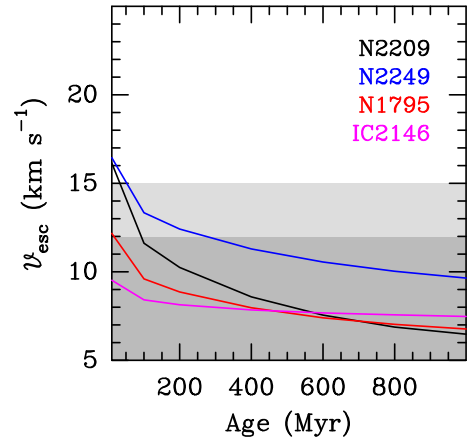}}
\caption{Escape velocities as a function of time for the four clusters in our
sample (identified in the legend at the top right). Light grey region represents
the critical escape velocity range, between \,12--\,15 \kms. The region below 12
\kms, in which cluster escape velocity is too low to permit the retention of the
material shed by the first generation and the presence of the eMSTO phenomenon,
is reported in dark grey.}   
\label{f:vel_time}
\end{figure}

In this context, it is important to verify that the clusters escape velocity is
above the derived threshold at the time the candidate polluters are present in
the clusters (i.e. at ages of $\sim$ \,5--\,30 Myr for massive stars and $\sim$
\,50--\,200 Myr for IM-AGB stars) and ejecting the material necessary for the
formation of the second generation. Fig.~\ref{f:vel_time} shows escape velocity
as a function of age, in a time range between 0 and 1 Gyr and with the
assumption on the level of initial mass segregation presented above. The
critical escape velocity range of \,12--\,15 \kms\ is depicted as the light grey
region in Fig.~\ref{f:vel_time}, while the region below 12 \kms, representing
the velocity range in which we do not observe eMSTOs in  intermediate-age star
clusters, is shown in dark grey. Note that NGC~2249 escape velocity remains
above 12 \kms\ for more than 250 Myr, while the escape velocity of NGC~2209
declines more rapidly and becomes lower than the threshold value after
$\approx$\,100 Myr. This is long enough for stars with ${\cal{M}} \ga 4 \;
M_{\odot}$ (i.e., all IM-AGB stars, see \citealt{vendan09}) to have lost all
their outer stellar envelope mass. Note also that given the very extended nature
of NGC~2209, the level of initial mass segregation experienced by the cluster
could be higher than the value we adopted in our analysis
(\citealt{mack+08b,goud+14}). In that case, NGC~2209's escape velocity would
remain above 12 \kms\ for  an even longer time. Conversely, NGC~1795 and
IC~2146, as expected, have an escape velocity below the threshold value at all
times. Hence, these results suggest that for clusters exhibiting an eMSTO, the
escape velocity remains above 12 \kms\ for a period of time long enough to
permit the retention of the material ejected by all the possible candidate
polluters, including the IM-AGB stars, which are the last in order of time to
start their  slow winds. Thus, it is worth to compare clusters escape velocities
with the wind speeds of the first generation candidate polluters: wind speeds
for FRMS range between ten to a few hundreds of \kms 
\citep{port96,porriv03,wuns+08}, massive binary stars have wind velocities in
the range 15\,--\,50 \kms\ (see e.g., \citealt{smit+02,smit+07}) and observed
wind velocities for IM-AGB stars in our Galaxy and the Magellanic Clouds are in
the range 12\,--\,18 \kms\ \citep{vaswoo93,zijl+96}. The latter range of wind
velocities is very similar to the ``critical'' escape velocity derived from our
dynamical analysis, suggesting that IM-AGB stars can be an important source of
the material needed for the formation of the second  generation. 

For what concerns scenarios that do not involve extended star formation, the
most popular is the ``stellar rotation'' scenario \citep{basdem09}. Using
theoretical arguments, the authors suggested that a distribution of rotation
velocities for stars between 1.2\,--\,1.7 $M_{\odot}$ can mimic the observed
morphology of the eMSTOs, shifting fast rotators stars to redder color in the
CMDs. However, \citet{gira+11}, using newly calculated evolutionary tracks for
non-rotating and rotating stars, reached opposite conclusions, that is stellar
models with rotation actually produce a modest blueshift in the CMD. Observations are needed to test these model predictions. To date, the only star cluster for which colors {\it and} rotation velocities have been determined for MSTO stars is the open cluster Trumpler~20 \citep{plat+12}. They showed that the distribution velocities of stars in Trumpler~20 includes the full range between 
0\,--\,180 \kms (which is similar to the range considered in the last models by \citealt{yang+13}), with a rather flat distribution, and that the fastest rotators are slightly \emph{blueshifted} (by $\Delta(V-I) = -0.01$) with respect to the slowest rotators. This is consistent with the prediction of \citet{gira+11}, and it suggests that the effect of stellar rotation on the morphology of MSTOs is significantly less pronounced that that advocated by \citet{basdem09}. 

Taking this into account, the results presented here suggest that the ``escape
velocity threshold scenario'' provides a valid explanation for the presence of
broadened MSTO region in intermediate-age star clusters and, in turn, that the
eMSTO phenomenon can be better explained by a range of stellar ages rather than
a range of stellar rotation velocities or interacting binaries.

\section{Summary and Conclusions}
We studied CMDs of two relatively low mass intermediate-age star clusters in the
LMC, namely NGC~2209 and NGC~2249, using new HST/WFC3 images. For comparison
purposes, we also re-analyzed archival HST/ACS images of NGC~1795 and IC~2146,
two other relatively low mass star clusters, for which the photometry was
already presented in \cite{milo+09}. We compared the CMDs of the clusters with
Monte Carlo simulations of SSPs in order to investigate the MSTO morphology  and
quantify the intrinsic widths of the clusters MSTO regions. To study the
physical and dynamical properties of the clusters, we derived their radial
surface number density distributions and we determined the evolution of the
clusters masses and escape velocities from an age of 10 Myr to  the current
age, considering models with and without initial mass segregation. The main
results of this paper can be summarized as follows: 
\begin{itemize}
\item NGC~2209 and NGC~2249 show a broadening of the MSTO region that can not be
  explained by photometric uncertainties, LMC field star contamination, or
  differential reddening effects. Comparison with Monte Carlo simulations of a
  SSP (including the effects of unresolved binary stars), show that the observed
  MSTOs are significantly wider than that of a single-age cluster with the same
  mass, [Z/H], and (average) age. On the other hand, NGC~1795 and IC~2146 show
  quite compact MSTOs, which are very well reproduced by their respective SSP
  simulations. 
\item NGC~2209 and NGC~2249 pseudo-age distribution are peaked towards younger
  ages and decline smoothly towards older ages. The lack of obvious secondary
  peaks in these distributions suggests that the morphology of the MSTO region
  of these clusters can be better explained by a spread in age than by two
  discrete SSPs.  
\item The physical properties of NGC~2209 and NGC~2249 are very different, even
  though their ages are almost identical: NGC~2209 is a very extended cluster
  (core radius $r_c = 6.64$ pc), while NGC~2249 is quite compact ($r_c = 2.46$
  pc). The presence of an eMSTO in both of these clusters suggests that the
  scenario proposed by K11, in which a cluster hosts an eMSTO only if it has a
  core radius larger than a certain value ($r_c \ga 3.7$ pc), can not fully
  explain the eMSTO phenomenon.  On the other hand, we find that the differences
  in MSTO properties among  all four clusters of our sample \emph{can} be
  explained by dynamical evolution arguments under the plausible assumption
  that   the currently extended clusters experienced stronger initial mass
  segregation than the currently compact ones. With this assumption, the derived
  early escape velocities for NGC~2209 and NGC~2249 are consistent with observed
  wind speeds for intermediate-mass AGB stars in our Galaxy and the Magellanic
  Clouds, which are considered to be one of the most probable candidate
  ``polluter'' stars of the first generation (the others being FRMS and massive
  binary stars). 
\item The non-ubiquity of eMSTOs among intermediate-age star clusters in the
  Magellanic Clouds along with the apparent dependence of the presence of eMSTOs
  on the initial dynamical properties of the clusters in the age range of
  10\,--\,100 Myr seems to indicate that age effects, rather than a range of
  stellar rotation velocities or interacting binaries, are responsible for the
  broadening of the MSTO region.  
\end{itemize}

\acknowledgments
Support for this project was provided by NASA through grant HST-GO-12908 from
the Space Telescope Science Institute, which is operated by the Association of 
Universities for Research in Astronomy, Inc., under NASA contract NAS5--26555.  
We made significant use of the SAO/NASA Astrophysics Data System during this
project.


\end{document}